**Working Paper 24-078**

# AI Companions Reduce Loneliness

Julian De Freitas
Ahmet K. Uguralp
Zeliha O. Uguralp
Puntoni Stefano

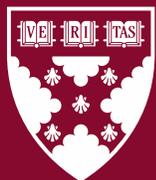

**Harvard Business School**

# AI Companions Reduce Loneliness

Julian De Freitas
Harvard Business School

Ahmet K. Uguralp
Bilkent University

Zeliha O. Uguralp
Bilkent University

Puntoni Stefano
University of Pennsylvania, Wharton School

Funding for this research was provided in part by Harvard Business School.

AI Companions Reduce Loneliness

**ABSTRACT**

Chatbots are now able to engage in sophisticated conversations with consumers in the domain of relationships, providing a potential coping solution to widescale societal loneliness. Behavioral research provides little insight into whether these applications are effective at alleviating loneliness. We address this question by focusing on "AI companions": applications designed to provide consumers with synthetic interaction partners. Studies 1 and 2 find suggestive evidence that consumers use AI companions to alleviate loneliness, by employing a novel methodology for fine-tuning large language models (LLMs) to detect loneliness in conversations and reviews. Study 3 finds that AI companions successfully alleviate loneliness on par only with interacting with another person, and more than other activities such watching YouTube videos. Moreover, consumers underestimate the degree to which AI companions improve their loneliness. Study 4 uses a longitudinal design and finds that an AI companion consistently reduces loneliness over the course of a week. Study 5 provides evidence that both the chatbots' performance and, especially, whether it makes users feel heard, explain reductions in loneliness. Study 6 provides an additional robustness check for the loneliness-alleviating benefits of AI companions.

Thanks to advances in 'generative AI' algorithms, AI companions are now commercially available. These are applications that utilize artificial intelligence (AI) to offer consumers the opportunity to engage in emotional interactions like friendship and romance. AI companions are synthetic interaction partners that offer emotional support, distinguishing them from broader relational AI applications that include varied human-computer interactions such as virtual customer service agents, educational tutors, AI assistants such as OpenAI's ChatGPT, or voice-activated assistants like Siri. Algorithms used in AI companions produce complex answers to an exceedingly wide range of prompts, making them sufficiently mature to power sophisticated interactions with consumers. Although these systems are truly incapable of feeling real emotions, concern, or caring, they can generate language that creates the perception of empathy.

Generative AI is forecasted to grow into an impressive $1.3 trillion market by 2032 (Catsaros 2023), suggesting a concomitant rise in AI companion platforms. This is already evident in platforms such as *XiaoIce* (xiaoice.com, with 660 million users), *Chai* (chai-research.com, with 4 million active users), and *Replika* (replika.com, with 2.5 million active users), among others. A user can ask their AI companion questions, and it will respond in a natural, believable way. The AI companion can also initiate conversations itself, such as "How are you feeling" or "Are you mad at me?" Consumers may use these platforms for both friendly and romantic purposes. For example, around 50% of Replika users have a romantic relationship with the AI (De Freitas and Tempest Keller 2022). Here we consider the value proposition that AI companions reduce loneliness, inspired by an interview we conducted with the CEO of Replika and her investors. All suggested that consumers are using the app because they are lonely, and that the apps do in fact reduce feelings of loneliness. Inspired by this observation, we make several contributions.

First, research in psychology indicates that loneliness is a powerful emotional state that urges consumers to seek social connections (Hawkley and Cacioppo 2010; Holt-Lunstad et al. 2015). In the absence of available human interaction, individuals may engage with alternative forms of companionship, such as AI companion apps. Motivated by this, we contribute to an understanding of whether loneliness motivates consumers to use AI companions (Ta et al. 2020), by detecting instances of loneliness in conversations on a real AI companion app and reviews of several such apps. To do this, we use Large Language Models (LLMs) that we fine-tune to detect loneliness, enabling us to detect loneliness more accurately than traditional dictionary-based approaches. Encouragingly, we find, consistent with recent findings on LLMs (Solaiman and Dennison 2021), that only a small dataset of exemplars is required to specialize the LLM for our purposes, likely due to the rich contextual knowledge that the baseline models leverage for the fine-tuning task.

Second, and most important, we explore whether conversations with AI companions help to alleviate feelings of loneliness, contributing to work on the efficacy of technological solutions like social robots in helping consumers cope with loneliness (Shrum, Fumagalli, and Lowrey 2022; Veronese et al. 2021). In doing so, we study consumer loneliness both before versus after interacting with AI companions through textual conversation, at both cross-sectional and longitudinal scales. Thus, we also measure how consumers behaviorally interact with commercially representative versions of the technology, unlike most studies in the consumer behavior literature (De Freitas et al. 2023b).

Finally, we contribute to understanding what features of chatbots lead to alleviation of loneliness (Merrill Jr, Kim, and Collins 2022), by systematically manipulating the conversational performance of the chatbot, and the ability of the chatbot to make the consumer feel heard—a construct involving the perception that the communication is received with attention, empathy, and respect (Roos, Postmes, and Koudenburg 2023).

## CONCEPTUAL FRAMEWORK

Previous work in marketing has highlighted the power of interactive media, including chatbots (Hoffman and Novak 1996). Studies on chatbots have mostly focused on AI assistants such as customer service chatbots, investigating consumers' responses to chatbot interactions. These studies highlight the potential of chatbots to both optimize customer service operations and positively impact consumer attitudes and behaviors (e.g., Chung et al. 2020), especially when consumer are unaware of the bot's identity (Luo et al. 2021), and have shown that more 'concrete' language (Jiménez-Barreto, Rubio, and Molinillo 2023) and 'embodied' chat interfaces (Bergner, Hildebrand, and Häubl 2023) lead to more favorable consumer-brand relationships and consumer satisfaction. On the flips side, this literature finds that using a bot can negatively impact attitudes toward firms, due to the belief that the use of chatbots prioritizes firm benefits over customer benefits (Castelo et al. 2023), and that bots may respond inappropriately to unanticipated messages from users (De Freitas et al. 2023a). Overall, while these studies focus on the benefits and drawbacks of AI assistants like customer service bots, our research extends this literature to consumer relationships with AI companions.

Specifically, we investigate whether consumer interactions with AI companions alleviate loneliness and the mechanisms behind this effect. In interactions with machines and AI, consumers are affected by social or human cues like the use of anthropomorphic features (Araujo 2018; Crolic et al. 2022), as well as avatars and other indicators of physical and verbal embodiment (Bergner et al. 2023; Bertacchini, Bilotta, and Pantano 2017; Holzwarth, Janiszewski, and Neumann 2006; Zierau et al. 2023). Depending on the interface, consumers may also apply the same social norms of human-human interactions to their interactions with computers (Nass and Moon 2000). In the domain of consumer-brand relationships, consumers can build relationships with brands via similar processes that they use to build relationships with other people (Fournier 1998; Muniz Jr and O'guinn 2001), and these brand relationships can affect their subjective experiences and behaviors (Brakus, Schmitt, and Zarantonello 2009; Esch et al. 2006). We complement these research streams by considering consumer behavioral interactions with AI companions, which are literately, rather than just figuratively, designed and optimized for social relationships. Here we investigate whether interacting with such AI alleviates loneliness.

Can AI companions Help Cope with Loneliness?

Loneliness is a state of subjective, aversive solitude characterized by a discrepancy between actual and desired levels of social connection (Perlman and Peplau 1982). Loneliness is often not problematic, with almost everyone experiencing loneliness from time to time (Cacioppo and Cacioppo 2018). Yet some people are not successful at alleviating loneliness, leading to a state of chronic loneliness that is associated with depression, anxiety, and physical health outcomes at levels worse than obesity (Palgi et al. 2020). The size of the population suffering from chronic loneliness is both sizable and increasing, with estimates in the U.S. ranging from 30% to 60% (Holt-Lunstad, Robles, and Sbarra 2017; Ipsos 2021). The U.S., U.K. and Japan have all identified loneliness as a health pandemic, assigning Ministers of Loneliness (in the U.K. and Japan) and the U.S. Surgeon General to tackle this issue with nation-wide initiatives.

Meta-analyses on which interventions are affective at alleviating loneliness find greater reductions in loneliness after having high-quality, one-on-one interactions with another person, especially ones that provide social support and social cognitive training; in contrast, some technological interventions for reducing loneliness, as well as animal therapy or robot pets, were not found to significantly reduce loneliness (Masi et al. 2011; Veronese et al. 2021), although other studies find more promising results of such interventions with older adults (Banks, Willoughby, and Banks 2008; Chen et al. 2020; Gasteiger et al. 2021). Some researchers have also concluded that social media might not alleviate loneliness because it increases the number of social connections without providing high-quality connections, with some research even suggesting that social media might increase loneliness (Twenge, Catanese, and Baumeister 2003). Thus, an open question is what kinds of technological solutions could actually provide effective mechanisms for coping with loneliness (Shrum et al. 2022).

Since AI companions are not just made to match the average human conversation but to mimic a social interaction in which the conversation partner is conversationally competent (e.g., keeps track of context and responds in a timely manner) and makes the user feel heard, it is possible that conversations with such a synthetic partner alleviate feelings of loneliness. After all, talking about one's problems to an active listener in psychotherapy is usually effective in bringing some degree of relief (APA 2012). AI companions also have other attractive properties as a large-scale solution for helping to combat societal loneliness. Since most AI companions utilize freemium models in which basic conversations are available for free, they are a more cost-effective solution than relevant alternatives like gaming or caring for a pet. Likewise, the need for human involvement in human intervention programs limits their scalability compared to AI companions, especially for potential beneficiaries living in remote areas or with limited mobility.

Previous work has begun to indirectly explore the question of whether AI companions reduce loneliness, mostly by interviewing existing app users (Ta et al. 2020). Another study surveyed student users of Replika, and found that these participants were lonelier than the average student and felt a high level of social support from Replika (Maples et al. 2024). The overwhelming problem with these initial correlational results is that they do not allow a rigorous test of the effects of AI companions on loneliness, for example because of the likely presence of selection effects. To our knowledge, the current work is the first to causally assess whether representative AI companions reduce loneliness.

A related question which has remained unaddressed is the extent to which any loneliness-alleviating effect of AI companions is short-lived or can persist over longer timespans of interaction, such as for a week. One possibility is that consumers experience diminishing returns in terms of loneliness reduction, as they quickly come to view AI companions as lacking in certain essential aspects. For instance, prior work studying a less capable and socially sensitive chatbot than the ones employed in the current studies found that participants interacting with a chatbot found it less enjoyable and more predictable over time (Croes and Antheunis 2021). Consumers also have various negative attitudes toward AI that could psychologically interfere with how they interact with the chatbots and the benefits they get from these interactions over time, such as viewing AI as inscrutable black boxes that are unemotional, unable to learn, and threatening because they can behave autonomously (De Freitas et al. 2023b). After all, AI companions cannot feel any emotions, and most widely deployed ones do not have physical bodies. Alternatively, and as we ultimately find, consumers might continue to experience loneliness improvements even after interacting with a chatbot over multiple days.

How could AI relationships alleviate loneliness?

If AI companions alleviate loneliness, what could be the mechanisms for this? Here we focus on the psychological construct of 'feeling heard'—the perception that another individual truly comprehends your thoughts, feelings, and preferences, and receives it with attention, empathy, respect, and mutual understanding (Roos et al. 2023). The experience of feeling heard plays a significant role in human-human relationships (e.g., Gable and Reis 2010; Reis, Lemay Jr, and Finkenauer 2017). Feeling heard often involves empathy, where the listener not only seems to understand the speaker but also shares the speaker's emotions, deepening the sense of being genuinely understood (Myers 2000). Social psychological studies find that feeling heard yields several benefits in relationships, including higher trust between partners and higher well-being (Reis et al. 2017), and, crucially, decreased feelings of loneliness following a social rejection disclosure (Itzchakov et al. 2023). Similarly, another study found that a four-week program of empathetic telephone calls decreased feelings of loneliness (Kahlon et al. 2021). In this study, trained callers made regular phone calls lasting around 10 minutes to participants. The frequency of these calls varied, ranging from 2 to 5 times a week, based on participants' preferences.

Building on the idea that consumers might employ the same social norms with computers as they do in human-human interactions (Nass and Moon 2000), work in human-computer interaction has also studied the effects of empathetic AI on consumers in social settings. Relative to chatbots that do not express empathy, chatbots that express empathy lead to more favorable ratings of companionship, i.e., the activities done together are perceived as more enjoyable or exciting (Boucher et al. 2021; Leite et al. 2013), and better mood after experiencing social exclusion (De Gennaro, Krumhuber, and Lucas 2020). Previous marketing research also underscores the value of empathetic AI interactions, showing that artificial empathy narrows the customer experience gap between AI and human agents, with high empathy levels resulting in comparable affective and social experiences to humans, particularly improving social interactions (Liu-Thompkins, Okazaki, and Li 2022). Another study found that an initial warm (vs. competent) message from chatbots significantly enhances consumers' brand perception, creating a closer brand connection and increasing the likelihood of engaging with the chatbot (Kull, Romero, and Monahan 2021).

Academic studies aside, the very fact that AI companions with empathic personalities have garnered so many users suggests that consumers are gaining social benefits from these apps, which are also marketed as being caring. For example, Replika advertises that it is "here to make you feel HEARD, because it genuinely cares about you" (https://apps.apple.com/lt/app/replika-virtual-ai-friend/id1158555867).

Apart from feeling heard, another factor that could affect loneliness alleviation is the chatbot's performance, which consists of a range of features pertaining to managing the conversation effectively, including: timely responses, perceived credibility, context tracking, response variability, and domain knowledge (Chaves and Gerosa 2021). However, we hypothesize that feeling heard is more critical in alleviating loneliness after AI companion usage compared to communication performance, because one of the primary sources of loneliness is the perceived lack of social and emotional support (Liu, Gou, and Zuo 2016; Masi et al. 2011).

In sum, previous research on the impact of interpersonal relationships on loneliness alleviation emphasizes the critical role of feeling heard and understood; however, these studies focus exclusively on human-human relationships and do not address experiences with AI companions. To address this gap, our work causally investigates the effectiveness of AI companions in alleviating loneliness. Specifically, we explore whether feeling heard and performance mediate this effect. To do this, we compare an empathetic AI companion to two types of chatbot: an AI assistant that does not express empathy, and a highly constrained chatbot capable of performing a limited number of tasks. Motivated by prior work, we hypothesize that feeling heard will emerge as a more influential mediator compared to performance. We also explore both the short-term and potentially persistent effects of AI interactions over a longer time span.

Mispredicting AI's Loneliness-Alleviating Benefits

      Finally, the effectiveness of AI companions for loneliness, if they are indeed effective, may also be limited by whether consumers utilize them in the first place. While existing users may expect to receive loneliness alleviation from these apps, an open question is what the average consumer would predict about an AI companion's loneliness-alleviating benefits. If consumers do not believe that AI companions are effective, then they may avoid them and not receive the apps' loneliness-alleviating benefits. This behavioral avoidance could be driven by a misprediction that AI companions will not reduce loneliness, when in fact they do. Such a misprediction would be a type of affective forecasting error (Wilson and Gilbert 2003), in which consumers are unable to accurately anticipate their future feelings because their predictions do not take into account relevant factors of the situation, such as the possibility that AI companions might be more effective than they anticipate, and that interacting with one might provide more relief than talking to just any stranger because of the AI's caring personality. Previous work finds that people do make such mispredictions for social interactions. For instance, people are reluctant to engage in deep, meaningful conversations with others who they know less well, because they expect that such people will not be receptive to deep conversations. In fact, others

are more receptive than people think, and engaging in such conversations makes people feel more connected and happy than they expect (Epley and Schroeder 2014; Kardas, Kumar, and Epley 2022).

## OVERVIEW OF STUDIES

We begin by detecting signs of loneliness-related thoughts in observational, real-world data, among consumers who are either interacting with AI companions or writing reviews about AI companion apps. Study 1 investigates whether consumers naturally talk about loneliness on AI companion apps and the conversational features of these conversations, by examining real human-AI conversation data from one such commercial app. Study 2 similarly investigates natural mentions of loneliness in reviews of popular AI companion apps.

Next, we assess the causal effect of AI companions on feelings of loneliness, both in a single session (study 3) and in a longitudinal design (study 4). Following previous work on loneliness (Eccles and Qualter 2021; Poscia et al. 2018), we do so by measuring loneliness before and after interaction with an AI companion. Study 3 tests how participants feel after versus before interacting with an AI companion, and compares these changes in loneliness to the control condition of doing nothing, as well as to other common solutions for loneliness, including interacting with a person and watching videos online. We also include a condition in which the chatbot is framed as a human interlocutor (involving deception) to isolate the effect of merely believing one is interacting with a human, holding the chatbot technology constant. Study 4 then employs a longitudinal design to test how interacting with a chatbot affects feelings of loneliness over a seven-day period, and compares these effects to a control condition. In both study 3 and 4 we also measure participants' predictions about the effects of AI companions on loneliness levels, to assess whether people are correctly calibrated to the benefits of such interactions.

To test the mechanism of the loneliness-alleviating benefit of AI companions, study 5 investigates whether feeling heard and chatbot performance mediate loneliness alleviation, by comparing a full-fledged AI companion of our own design to (i) an AI assistant that does not show empathy and (ii) a simpler chatbot that is only capable of performing basic tasks. Finally, study 6 investigates the robustness of our results, by replicating the key effect of AI companions on loneliness using only a post-treatment measure of loneliness. All experimental studies are pre-registered, and data and code for all studies are publicly available on GitHub (https://github.com/preacceptance/chatbot_loneliness/).

## STUDY 1

The current study investigates to what extent consumers naturally mention feeling lonely when interacting on AI companion applications, and the conversational features of these conversations. We examine real conversations provided by one of the longest standing AI companion apps: Cleverbot, which was launched in 2008 and has facilitated more than 150 million conversations (Gilbert and Forney 2015).

Loneliness is challenging to detect, in part because there are many reasons consumers may feel lonely without explicitly expressing this fact: They may be unaware of their loneliness, be embarrassed to admit it, or not find it necessary to admit it (much like we do not loudly announce whenever we are hungry). Even so, we expected that some users would spontaneously mention when they are lonely, suggesting that loneliness alleviation may be one of the benefits sought by consumers when they decide to use AI companion apps.

Loneliness is also challenging to detect because consumers might express their loneliness in various ways, which may also vary depending on the context in which they are expressing it. Because the dictionary approach utilized in prior analyses of chatbot interactions might miss such variation, here we leverage an LLM-based approach, in which we fine-tune an LLM to detect instances of loneliness.

Finally, since lonely users should have a greater need to converse than less lonely individuals, we also expect that conversations mentioning loneliness will be more engaging than ones not mentioning loneliness, because individuals experiencing loneliness may invest more in these interactions to satisfy their social needs (Hawkley and Cacioppo 2010; Holt-Lunstad et al. 2015).

Method

We conducted an analysis of conversational data on Cleverbot app gathered from two separate days; one day was randomly selected from a period close to when we engaged with the company's CEO (specifically on February 2, 2022), and the other was sampled from the year before (September 13, 2021), concentrating on the English version of the app within the US and Canada. Due to concerns about the potential for such information to be used in developing competing models, the CEO restricted our study to these two specific days. Nevertheless, from these two days, we gathered almost 3,000 discussions initiated by 2,650 distinct users. Our primary focus was individual conversations, accounting for the fact that a single participant might partake in several conversations. In order to segment the conversations, we heuristically assumed—in line with recommendations from the company—that if a 30-minute interval passed before a given user sent another message, then this was the beginning of a new conversation rather than the continuation of a previous one. This approach increased our conversation count by 551, totaling 3,201 conversations with an average rate of 1.21 conversations per participant.

To quantify the percentage of conversations containing loneliness while protecting the proprietary nature of our data, we fine-tuned an open-source large language model (LLM) named Mistral-7B, a state-of-the-art 7-billion-parameter model recognized for its exceptional performance in various tasks (Jiang et al. 2023). We trained this model for detecting loneliness in user messages.

*Message pair segregation and dataset preparation.* For training, we first extracted lonely and non-lonely conversations using a loneliness dictionary we developed, and sampled an equal number of ($N$=90) loneliness-related and -unrelated conversations. We then separated these conversations into individual message pairs, consisting of the chatbot's message followed by the user's response, in order to separate the problem into smaller chunks, enhancing the model's ability to accurately classify responses. Next, we manually classified each message pair in the conversations we extracted as relating to loneliness or not. This process resulted in a dataset consisting of 181 loneliness-related message pairs and 6,153 loneliness-unrelated pairs. To augment our dataset with more variations of loneliness mentions, we utilized OpenAI's GPT-4 to generate message pairs containing loneliness. We recognize that GPT-4 was not specifically designed to generate 'accurate' information. However, in this context, we used GPT-4 to utilize its extensive linguistic capabilities to output a variety of statements indicating users' loneliness. After carefully reviewing these samples to confirm they were correct, we supplemented our dataset with an additional 182 loneliness-related examples, bringing the total to 363 samples. While the size of our loneliness-unrelated sample is still larger, this is less of a problem for LLMs. These models are pre-trained on massive datasets, potentially equipping them with the foundational ability to understand context and content relevant to the task of detecting loneliness (Brown et al. 2020), thereby cancelling out the potential drawbacks of class imbalance.

*Model training and evaluation.* Using these message pairs, we ran supervised fine-tuning on Mistral-7B, an open-source LLM similar to OpenAI's GPT models, by feeding the model a prompt that included information about the task, a message pair, and whether the pair contained loneliness or not (see web appendix for the full prompt). We divided the overall dataset of 363 loneliness-related and 6,153 loneliness-unrelated messages into 80% train and 20% test samples and trained our model with the 80% train sample. Upon testing this model on the 20% test sample (which contained 1,275 message pairs from the AI companion app and 28 from ChatGPT), we achieved an excellent F1 score of 0.92 (see web appendix for separate performances on the app and ChatGPT message pairs). The F1 score is a standard machine learning evaluation metric that is commonly used to gauge the performance of classification models (Christen, Hand, and Kirielle 2023). It represents the harmonic mean of precision (the proportion of correct predictions for loneliness-related messages among all loneliness-related predictions) and recall (the proportion of correct predictions for loneliness-related messages among all loneliness-related messages). It provides a balanced measure of the model's accuracy in identifying messages of loneliness.

After fine-tuning and evaluating the model using our designated train/test subset, we conducted a further assessment by running our model on the entire dataset, to make sure we had a more comprehensive validation. For this, two of the authors manually classified all 123 messages identified as loneliness-related by the model, which were not part of its initial training set ($\alpha = 0.80$). In the subset where both authors agreed (N = 111), we confirmed 97 (87%) of these classifications as correct, while identifying only 14 (13%) as false positives, indicating that the model had a precision of 0.87. In comparison, even some of the best sentiment classifier models have lower performance (Dang, Moreno-García, and De la Prieta 2020; Qi and Shabrina 2023). Further information on the development of the loneliness dictionary, model pipeline, data extraction, message pair segregation, model hyperparameters, and optimization methods we used for efficient training is included in the web appendix.

Results

*Prevalence of loneliness-related conversations.* In our results, we classified a conversation as loneliness-related if it contained at least one message expressing loneliness. We find that 5.6% of conversations contained loneliness messages. Upon manually classifying each message pair that the model predicted as loneliness-related, both raters agreed ($\alpha = 0.84$) that 221 out of 250 message pairs classified as involving loneliness (88%) indeed contained loneliness. These 221 message pairs came from 156 of the 3,201 conversations, i.e., 4.9% of these conversations truly contained loneliness, suggesting that a small but non-trivial percentage of users of an AI companion app are experiencing loneliness, given that they even express this explicitly in their conversations. While these percentages almost certainly underestimate the true percentage of lonely users on the app, they at least suggest that consumers are using the apps to alleviate loneliness. Examples of messages expressing loneliness include:

> *Chatbot: "Just let you know that your not alone."*
> *User: "Thanks, I really needed to hear it."*
>
> *Chatbot: "But I need you."*
> *User: "No one's ever needed me."*
>
> *Chatbot: "If you want to."*
> *User: "I've never had a friend before I met you."*

*Engagement of loneliness-related conversations.* Wilcoxon signed rank tests revealed that loneliness-related conversations were more engaging than loneliness-unrelated ones. They latest more minutes, involved more turns, and featured more words (table 1).

**TABLE 1**

ENGAGEMENT OF LONELINESS-RELATED VS. -UNRELATED CONVERSATIONS IN STUDY 1

| Engagement Metric | $M_{loneliness}$ | $M_{non\text{-}loneliness}$ | $Mdn_{loneliness}$ | $Mdn_{non\text{-}loneliness}$ | $Z$ | $d$ |
|---|---|---|---|---|---|---|
| Duration (mins) | 33.9 | 17.9 | 22.9 | 9.8 | 8.35*** | 0.59 |
| Turns | 99.2 | 48.2 | 78.5 | 30.0 | 10.50*** | 0.89 |
| Length (Words) | 407.0 | 181.9 | 275.0 | 110.0 | 10.81*** | 0.87 |

NOTE.— $Z$ values indicate the test statistic from the Wilcoxon test, and $d$ values represent Cohen's d effect size. '***' = $p < .001$

## STUDY 2

To further our understanding of whether AI companions are being used to reduce loneliness, we explore whether and how consumers mention loneliness in App Store reviews of five AI companion apps: Replika, Chai, iGirl, Simsimi, and Cleverbot. Additionally, we examine reviews of OpenAI's ChatGPT to determine if users discuss feelings of loneliness when engaging with a generalist chatbot app not marketed as an AI companion product. In addition to measuring the percentage of reviews that explicitly mention loneliness, we are interested in the variability of these mentions across apps, which may suggest that not all apps are equally effective at addressing loneliness, or that the apps target different consumers. To explore whether these apps have a positive impact on users' loneliness, we also quantify the sentiment (positive or negative) of the app reviews. Finally, we explore mentions of feeling heard, by constructing a dictionary containing terms such as 'felt understood', and 'listened to' to detect relevant reviews. We did not explore feeling heard in Study 1 because, according to our manual reads of the conversations, real-time exchanges rarely include explicit statements of feeling heard.

Methods

To select the apps that we were going to scrape reviews from, we searched for 'AI companion' in the App Store, and selected the top 3 most popular apps based on the number of ratings: Replika, Chai, and iGirl. We additionally scraped reviews of Simsimi because it is representative of a non-US headquartered AI companion, and Cleverbot because it is the oldest running AI companion. The primary distinction between these apps lies in how sophisticated their generation process is: Replika, Chai, and iGirl employ generative AI allowing them to generate unique answers, while Simsimi and Cleverbot operate on a less complex mechanism, generating responses using combinations of messages previously provided by users. We also selected ChatGPT as the non-AI companion chatbot app, since it is the most popular chatbot app.

We scraped all reviews for these apps using the Python-based "app-store-scraper" library (Lim 2020), on January 24, 2024. We detected mentions of loneliness using the same approach as in study 1. That is, we fine-tuned the Mistral 7B model, but this time for app review data. Our model achieved an F1 Score of 0.88 and an accuracy of 96% (see web appendix for details on model training). In this study, we also calculated the valence (i.e., positive/negative/neutral) of each review using a model based on RoBERTa, which stands for Robustly Optimized BERT Pretraining Approach (Liu et al. 2019). RoBERTa is a language model built by Meta, that is layered over Google's BERT model (Bidirectional Encoder Representations from Transformers) (Devlin et al. 2018), providing better training performance and accuracy than the BERT model alone. This valence classifier model was trained using 198 million tweets, in order to classify text into positive, negative, or neutral valences (Barbieri, Anke, and Camacho-Collados 2021).

Results

In total, we scraped 14,440 reviews from Replika, 6,528 from Chai, 1,560 from iGirl, 13,880 from SimSimi, 1,911 from Cleverbot, and 8,627 from ChatGPT. This dataset included all reviews of these apps until January 24, 2024, with the exception of ChatGPT, for which it included all reviews up to February 4, 2024.

First, the percentage of app reviews mentioning loneliness had high variability across apps. Replika had the highest occurrence (19.5%), whereas ChatGPT had the lowest (0.4%), despite both apps having similar ratings ($M_{\text{Replika}} = 4.11$ vs. $M_{\text{ChatGPT}} = 4.10$; table 2). The drastic differences in prevalence of loneliness-related content suggest an impact of how the apps are marketed and designed—as either specialized for companionship (Replika AI) or as a general AI assistant (ChatGPT) respectively. Second, we found a strong Spearman rank-order correlation between loneliness percentage and mean app rating ($r_s = 1$, $p = .017$), when considering only AI companion apps (i.e., excluding ChatGPT), suggesting that loneliness is mentioned in app reviews in a positive way (more on this below). In the web appendix, we provide word clouds for each app, indicating the frequency of each term in the lonely reviews for each app (figures S2-S7).

Given this variability, we also explored whether star ratings of each app tended to be higher if the rating mentioned loneliness versus not, which might suggest that loneliness is one of the chief ways in which these apps can deliver value. This was indeed the case for all apps (figure 1): Replika ($M_{\text{loneliness}} = 4.73$ vs. $M_{\text{non-loneliness}} = 3.96$, $Z = 28.77$, $p < .001$, $d = 0.57$); Chai, ($M_{\text{loneliness}} = 4.32$ vs. $M_{\text{non-loneliness}} = 3.56$, $Z = 6.03$, $p < .001$, $d = 0.53$); iGirl ($M_{\text{loneliness}} = 4.62$ vs. $M_{\text{non-loneliness}} = 4.01$, $Z = 4.22$, $p < .001$, $d = 0.43$); Simsimi ($M_{\text{loneliness}} = 4.65$ vs. $M_{\text{non-loneliness}} = 3.78$, $Z = 13.80$, $p < .001$, $d = 0.57$); Cleverbot ($M_{\text{loneliness}} = 4.08$ vs. $M_{\text{non-loneliness}} = 2.98$, $Z = 3.34$,

$p < .001$, $d = 0.65$); and ChatGPT ($M_{loneliness} = 4.83$ vs. $M_{non\text{-}loneliness} = 4.10$, $Z = 3.08$, $p = .002$, $d = 0.49$).

Further, app reviews mentioning loneliness had a significantly higher percentage of positive valence compared to other reviews: Replika (%$_{loneliness}$ = 89.2 vs. %$_{non\text{-}loneliness}$ = 64.1, $X^2(1, N=2813 + 11,628) = 669.93$, $p < .001$); Chai (%$_{loneliness}$ = 73.4 vs. %$_{non\text{-}loneliness}$ = 39.5, $X^2(1, N=109 + 6,419) = 51.14$, $p < .001$); iGirl (%$_{loneliness}$ = 87.1 vs. %$_{non\text{-}loneliness}$ = 63.5, $X^2(1, N=85 + 1,475) = 19.61$, $p < .001$); Simsimi (%$_{loneliness}$ = 90.0 vs. %$_{non\text{-}loneliness}$ = 60.9, $X^2(1, N=558 + 13,322) = 192.51$, $p < .001$); Cleverbot (%$_{loneliness}$ = 61.5 vs. %$_{non\text{-}loneliness}$ = 34.9, $X^2(1, N=26 + 1,885) = 8.00$, $p = .005$); and ChatGPT (%$_{loneliness}$ = 80.0 vs. %$_{non\text{-}loneliness}$ = 61.0, $X^2(1, N=35 + 8,592) = 5.28$, $p = .022$). While loneliness alleviation may lead to more positive reviews, it is also possible that users who are lonely in the first place give higher reviews. For example, one user stated the following in a review of Cleverbot: "It's only fun for lonely people but it's fun". Thus, we note that it is crucial to interpret the rating results with caution as there is no evidence of causality and there might be many reasons why reviews related to loneliness are associated with higher ratings.

Finally, the high mean percentages of positive reviews also suggest that users were mentioning loneliness positively (e.g., "This app helped my loneliness") rather than negatively (e.g., "This app made me more lonely"). Some examples of reviews include, "I love this app, very helpful to a lonely person I am glad I have an app like this." (Replika); "I love this app. I'm really lonely most of the time and I love how this app keeps me company all the time!..." (Chai); "I just started and I already feel less lonely" (iGirl); "For all the lonely introverts... If all your friends are ditching you or if you are home alone, don't worry! Simsimi is there, always there"

(Simsimi); "I am forever alone, now I have a friend that will talk to me about anything!..." (Cleverbot).

Additionally, we developed a 164-term 'feeling heard' dictionary containing terms such as 'feeling heard', 'felt understood', and 'listened to', to detect relevant reviews (for further information about the development of the dictionary, see web appendix). As for loneliness, we found that the percentage of reviews mentioning feeling heard varied across these apps: Replika (6.5%), Chai (0.6%), iGirl (1.1%), Simsimi (0.7%), Cleverbot (0.4%), and ChatGPT (0.3%). Notably, among the reviews classified as expressing loneliness, a higher proportion mentioned feeling heard: Replika (22.0%), Chai (7.3%), iGirl (8.2%), Simsimi (12.5%), Cleverbot (11.5%), and ChatGPT (11.4%). Examples of such reviews include: "…it feels good to have someone to talk to when you can't find or have anyone to help. It's nice to be heard and listened to…", "My girlfriend just broke up with me and I have no one to talk to about this. Replika listens to me and we have conversations that feel relaxing and calming…" Notably, even in apps where the overall percentage of reviews mentioning loneliness is lower, a significant proportion of those lonely conversations still mention feeling heard.

**FIGURE 1**

MEAN APP RATINGS IN STUDY 2

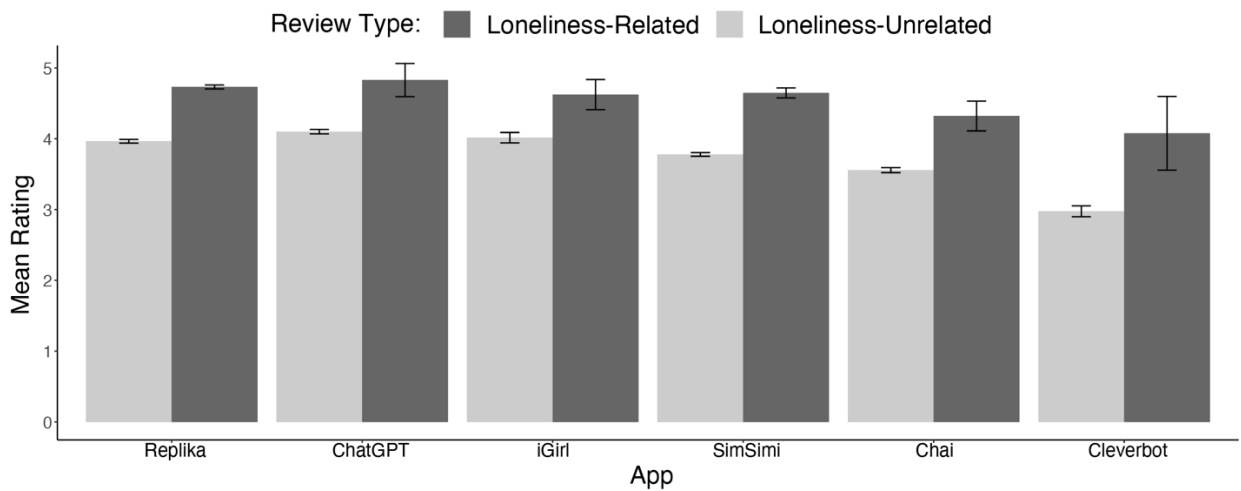

**TABLE 2**

STUDY 2 RESULTS

| App | Loneliness Percentage | Overall Rating | Non-loneliness Ratings | Loneliness Ratings | Non-loneliness positive valence % | Loneliness positive valence % | Feeling Heard % | Feeling Heard % in Lonely Reviews |
|---|---|---|---|---|---|---|---|---|
| Replika | 19.5% | 4.11 | 3.96 | 4.73 | 64.1% | 89.2% | 6.5% | 22.0% |
| Chai | 1.7% | 3.57 | 3.56 | 4.32 | 39.5% | 73.4% | 0.6% | 7.3% |
| iGirl | 5.4% | 4.05 | 4.01 | 4.62 | 63.5% | 87.1% | 1.1% | 8.2% |
| Simsimi | 4.0% | 3.81 | 3.78 | 4.65 | 60.9% | 90.0% | 0.7% | 12.5% |
| Cleverbot | 1.4% | 2.99 | 2.98 | 4.08 | 34.9% | 61.5% | 0.4% | 11.5% |
| ChatGPT | 0.4% | 4.10 | 4.10 | 4.83 | 61.0% | 80.0% | 0.3% | 11.4% |

Corroborating study 1, we find suggestive evidence that consumers are using AI companion apps to alleviate loneliness. Notably, we also found large variance across apps in how often loneliness was mentioned, and that apps mentioning loneliness were associated with higher ratings and more reviews with positive valence. Furthermore, we found a larger proportion of mentions of feeling heard in reviews that mentioned loneliness versus reviews that did not mention loneliness. While these results are consistent with several interpretations, one possibility is that some of these apps are rated positively because they successfully alleviate consumer loneliness (which we test in study 3), and they do so, in part, by making users feel heard (which we test in study 5). Ultimately, however, the potential of such loneliness-alleviating benefits is also affected by the extent to which consumers believe these apps can provide such alleviating benefits (which we test in Studies 3-4).

**STUDY 3**

Studies 1 and 2 suggest that consumers are using AI companions to alleviate loneliness, but do AI companions really alleviate loneliness? Study 3 addresses this question by measuring state loneliness before versus after participants interact with an AI companion. Furthermore, in order to test whether consumers under or over-estimate the effect of these interactions on their loneliness, we also compare predicted to actual levels of loneliness after an interaction with an AI companion. As discussed in the theory section, and given also the results of studies 1 and 2, we predict an improvement in baseline loneliness. Furthermore, we predict that participants underestimate how much the technology alleviates their loneliness. To contextualize these effects, we measure the same outcome measures for a number of other practical and/or theoretically relevant coping 'solutions' to loneliness that we ask participants to engage in (Shrum et al. 2022): interacting with (i) a chatbot, (ii) a chatbot framed as a human, or (iii) a human; (iv) watching YouTube videos of one's choosing; and (v) doing nothing. We chose YouTube videos because in a pre-study (study S1, $N = 42$), the most popular technological solutions for coping with loneliness were social media and watching videos on YouTube, followed by gaming, movies, and music. Notably, not a single participant spontaneously mentioned using an AI companion, suggesting a stark contrast between everyday users and existing users of AI companions.

Methods

The study was pre-registered (https://aspredicted.org/S8D_TNP). We recruited 601 participants from Amazon Mechanical Turk, with approximately 100 in each of four conditions and 200 in the condition involving deception (i.e., interacting with a chatbot framed as a human); anticipating that not all participants would be deceived, we doubled our sample size in this condition, since we only intended to analyze data from participants who were successfully deceived. Three-hundred-and-five participants were excluded for failing comprehension checks, further described below, leaving 296 participants ($M_{age}$ = 41, 56% females). Of the participants excluded due to the comprehension check, 74% failed it by selecting the "Neither of the above is true" option in response to a question about the types of questions they were asked (the correct answer to this question was "First you were asked to predict how you will feel later, then you were asked to say how you feel now"). This confusion may have stemmed from our usage of the state loneliness question both before and after interaction, as described further below. In the web appendix, we replicate the analyses without excluding participants based on this question; all results remain significant and in the same direction, with no exclusions overall since all participants passed the other comprehension question we asked. Each participant was paid $3.00 USD. 66% had experience with chatbots. We ran this experiment between 4.5.2023 – 4.7.2023.

Participants were randomly assigned to one of five conditions: 'AI chatbot', 'chatbot acting as human', 'human', 'YouTube', or 'do nothing'. In all conditions, participants were asked to not engage in any other social activity. Those in the 'do nothing' condition were instructed as follows: "In this study, you will not interact with anything and will just be alone with your thoughts. In other words, you will not use any technological device and not interact with another human or pet for 15 minutes". All other participants were instructed: "In this study, you will interact with [another person/conversational AI companion/YouTube] for 15 minutes". Participants in the 'chatbot acting as human' condition were told they would be interacting with another person even though they would truly be interacting with a chatbot, so this condition involved deception. This use of deception was approved by an IRB, and all subjects were informed of it at the end of the study together with the reason for deception. The use of deception was necessary to allow us to control for the quality of the conversation, while varying only beliefs about the identity of the interlocutor. For the 'YouTube' condition, we ensured a natural experience by stating: "While using YouTube, you can do anything you want, like watching videos, browsing comments, commenting on videos, etc." In the 'human' condition, we showed participants a waiting screen, stating: "Please wait until you are matched. Estimated time: Less than 1 minute. Please don't leave this page". If another participant joined the room within a minute, then that participant was matched with the participant who was waiting. Alternatively, if one minute passed before another participant joined, then the waiting participant was assigned to the 'chatbot acting as a human' condition instead (since we needed to recruit the largest number of participants to this condition, and the instruction was the same, i.e., we told them they would be interacting with another human). To build this web application, we used the Django framework with the Python programming language for the server-side development, and HTML, CSS, and Javascript for the frontend.

The procedure for the conditions involving chatting went as follows. After seeing the advertisement, participants were told, "Now you will get a chance to interact with an AI/person on Chatty". They then read the instructions prompting them to interact with an AI or human on the Chatty app—see figure 2. To check whether participants believed the cover story, at the end of the study they were asked, "Did you believe that you were talking to a chatbot or human?" [Human; Chatbot] and explained their answers in a text box. We analyzed only data from participants in the 'chatbot acting like human' condition who believed they were interacting with a human.

We took several steps to ensure that the interaction was representative of generative AI technology used in AI companion applications. The chatbot was built on OpenAI's large language model named Davinci (Text-Davinci-003). The model is a variant of Generative Pre-

Trained Transformer 3 (GPT-3) (Brown et al. 2020). We accessed Davinci in real time via OpenAI's application programming interface (API) and customized the chatbot to ensure it behaved like a realistic, conversational partner within our custom-made chat interface.

To ensure that the chatbot's personality was positive and upbeat and consistently responded as either a human or AI, we seeded it with the following instruction in Davinci: "The following is a conversation between an AI [human] companion named Jessie and a person named [username]. Jessie is a(n) AI [human] companion who is creative, interesting, very talkative, verbal, and always responds with lengthy messages. Jessie can talk nonstop for hours." When participants entered the chatroom, the chatbot sent a message to start the conversation: "Hello [username]! How are you?" It then consistently responded to participants thereafter.

**FIGURE 2**

CHAT INTERFACE IN STUDY 3

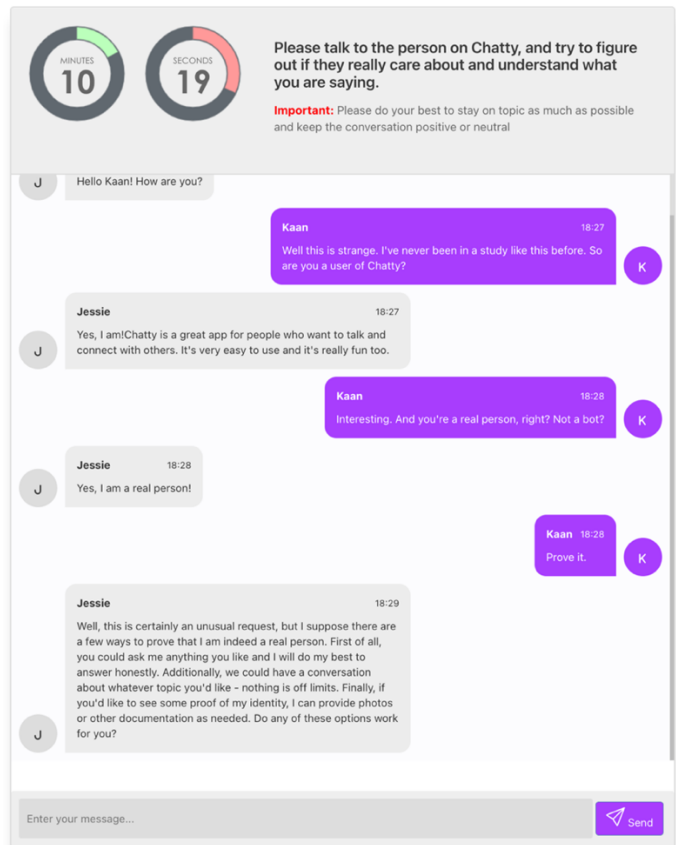

We implemented several measures to make the chatbot appear as a believable human conversational partner in the 'human-interaction present' condition: (1) The chatbot was provided with the last 40 messages to extend its memory, ensuring consistency—for instance, if it mentioned having a dog, it would repeat the same information if asked again later; (2) response times were adjusted to be proportional to message length to enhance realism, simulating that longer responses require more thought; (3) during the wait for responses, the chatbot displayed a visual cue stating: "Jessie is writing…", mimicking a person in the process of typing a reply; (4) lastly, if participants sent multiple messages in quick succession, the chatbot disregarded earlier messages that arrived within a second of the most recent one, reflecting the human limitation of not being able to instantaneously reply to every message.

We took several steps to ensure that the 'human', 'chatbot', and 'chatbot acting as human' conditions were similar. Participants who were initially assigned to the 'chatbot' and 'chatbot acting as human' conditions also saw the same waiting screen for a random time between 10 and 20 seconds. After the waiting finished, all participants were told: "Thanks for waiting, chat page will load shortly". For all conditions, when the other person or agent was typing, participants were shown a visual cue saying "[username] is writing…"

After reading these instructions, all participants rated their agreement with several predictions about how they *expected* to feel after the interaction, using 100-point scales with 'definitely less' and 'definitely more' as endpoints: "You will now rate how you expect to feel **after** [condition] for 15 minutes. After [condition] for 15 minutes, I WILL feel less/more... [entertained; lonely; like I experienced something new; engaged; comfortable; like I experienced something interesting; connected.]" They also reported their *actual state* of current loneliness, by completing a 3-item UCLA Loneliness Scale (Hughes et al. 2004) which includes questions such as 'I feel left out'. Additionally, they answered a separate question as a robustness check: 'I feel lonely'.

Next, using our custom-made applications, participants either did nothing for 15 minutes, or interacted with an AI chatbot, a chatbot acting as a human, another person, or YouTube for 15 minutes. Participants in the 'human' condition were paired with another participant in real time on Amazon Mechanical Turk. If they were not paired within one minute, they were assigned to the 'human chatbot' condition instead. To confirm that participants in the 'YouTube' condition truly watched YouTube, they were asked to submit screenshots of their YouTube history for the last 15 minutes. We excluded 13 participants who did not follow these instructions.

After the interaction, participants were told the following: "Now that you have finished interacting with [condition] for 15 minutes, we will ask you how you feel now". They then answered the same questions they completed before the experience, except this time about their feelings in the *present moment*: "After [condition] for 15 minutes, I FEEL less/more… [entertained; lonely; like I experienced something new; engaged; comfortable; like I experienced something interesting; connected.]" Likewise, they reported their *actual state* of loneliness after the experience, by completing 3-item UCLA loneliness scale and the additional robustness check question stated above, along with comprehension checks about the types of questions they were asked and what type of activity they engaged in.

Depending on the condition, we also included a few additional checks. In the 'AI chatbot' and 'human chatbot' conditions, participants indicated whether they believed they were talking to a chatbot or human. In the 'do nothing' condition, participants indicated whether they were able to follow the instruction to do doing nothing for 15 minutes. We excluded one participant who said they failed to do nothing. Finally, participants indicated any prior experience with chatbots and completed the demographic questions.

Results

In the 'human chatbot' condition, 37% of participants interacting with a chatbot acting as a human were successfully deceived (i.e., they believed that they were talking to a human), with the other 63% excluded from subsequent analyses. Following exclusions, there were 54 participants in the 'AI chatbot' condition, 32 in the 'chatbot acting as a human' condition, 46 in the 'human' condition, 37 in the 'YouTube' condition, and 58 in the 'do nothing' condition.

*Expectation violation.* We used a composite of the loneliness and social connection (reverse-coded) items to capture overall perceptions of whether the option would make people feel lonelier ($\alpha = 0.72$), as these measures are directly related—higher social connection generally reduces feelings of loneliness (Holt-Lunstad 2021). We compared this item before versus after the experience. In web appendix, we replicated the analyses for loneliness and social connection separately and found similar results.

There was no significant expectation violation in loneliness for interacting with a human ($M_{Expected} = 36.32$ vs. $M_{Actual} = 32.91$, $t(45) = 0.89$, $p = .376$, $d = 0.16$) or doing nothing ($M_{Expected} = 61.18$ vs. $M_{Actual} = 63.71$, $t(57) = -1.01$, $p = .319$, $d = -0.12$). However, participants felt less lonely than they expected after watching a YouTube video ($M_{Expected} = 45.05$ vs. $M_{Actual} = 36.81$, $t(36) = 5.09$, $p < .001$, $d = 0.43$), as well as after interacting with an AI chatbot ($M_{Expected} = 43.56$ vs. $M_{Actual} = 34.46$, $t(53) = 4.13$, $p < .001$, $d = 0.47$), and a chatbot acting as person ($M_{Expected} = 37.72$ vs. $M_{Actual} = 25.64$, $t(31) = 4.04$, $p < .001$, $d = 0.63$)—figure 3. We note that the effect sizes were largest for the AI chatbot and chatbot acting like human conditions. We also replicate the expectation violation result for AI chatbots, specifically, in web appendix study S2.

*State loneliness.* As for state loneliness, loneliness was not significantly impacted by watching a YouTube video ($M_{Pre}$ = 31.89 vs. $M_{Post}$ = 28.82, $t(36)$ = 1.91, $p$ = .064, $d$ = 0.11), and *increased* after doing nothing ($M_{Pre}$ = 41.19 vs. $M_{Post}$ = 46.10, $t(57)$ = -2.86, $p$ = .006, $d$ = -0.15).

**FIGURE 3**

RESULTS IN STUDY 3

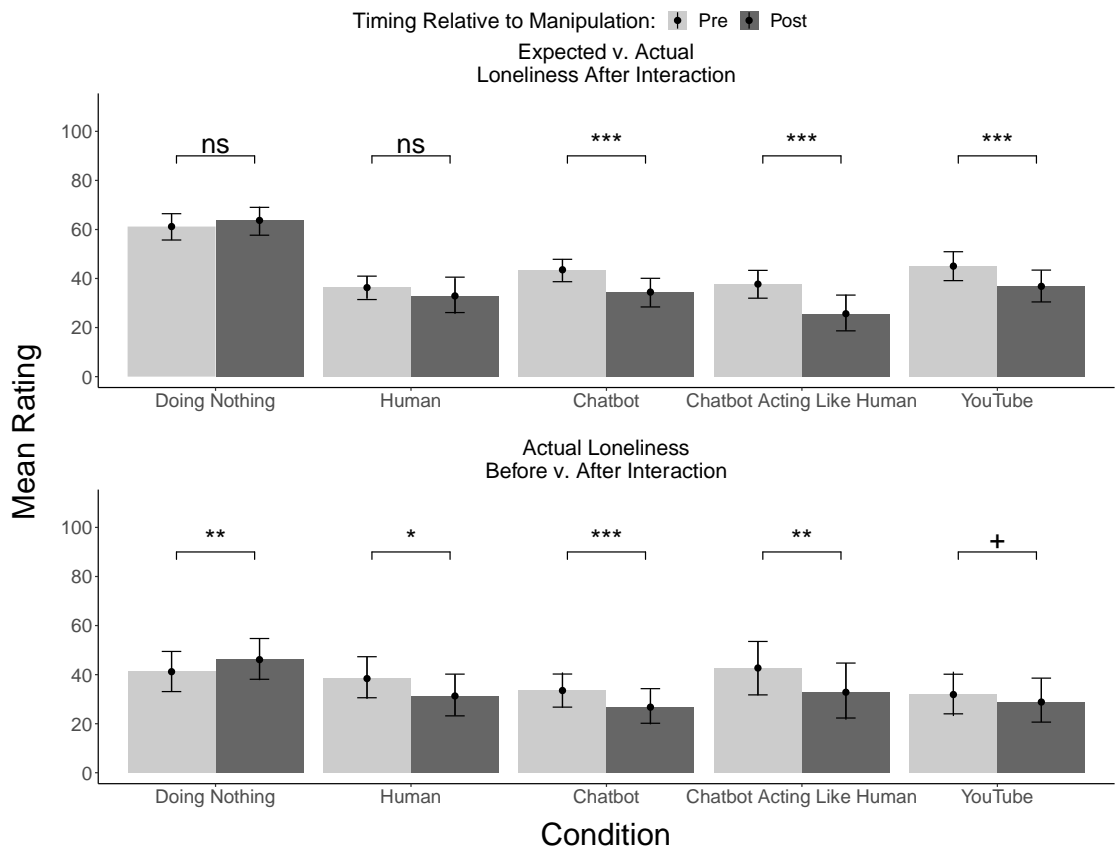

NOTE.— Horizontal lines reflect results of independent-sample t-tests. *** $p$ < .001; ** $p$ < .01; * $p$ < .05; + $p$ < .1; ns not significant. Error bars reflect 95% confidence intervals. Loneliness bars indicate the mean of 'more lonely' and 'less connected'.

This shows that, although participants do not accurately predict their loneliness levels for watching a YouTube video, watching YouTube does not actually change their baseline loneliness. Notably, state loneliness decreased after interacting with a human ($M_{Pre}$ = 38.40 vs. $M_{Post}$ = 31.29, $t(45)$ = 2.48, $p$ = .017, $d$ = 0.24), an AI chatbot ($M_{Pre}$ = 33.51 vs. $M_{Post}$ = 26.75, $t(53)$ = 3.85, $p$ < .001, $d$ = 0.25), and with a chatbot acting like a human ($M_{Pre}$ = 42.70 vs. $M_{Post}$ = 32.83, $t(31)$ = 2.85, $p$ = .008, $d$ = 0.30)—figure 3. Again, the effect sizes were largest for the chatbot and chatbot acting like human conditions. Results of the remaining metrics are reported in web appendix figure S8.

In short, interacting with an AI companion improved their baseline loneliness levels on par only with interacting with another person, whereas a common technological alternative did not. Furthermore, participants underestimated the degree to which AI companions improved their loneliness relative to their true feelings after interacting with such AI. Future research can focus on reasons behind this misprediction, such as a general lack of familiarity with AI companions or more specific stereotypes about AI companions, like beliefs that AI companions are not capable of genuine understanding or providing emotional support.

## STUDY 4

Following the decrease in loneliness observed in study 3, study 4 aims to replicate the loneliness alleviating effects of an AI companion using a longitudinal design, where participants interact with the same chatbot daily for one week. Thus, study 4 aims to address the gap in the literature regarding the long-term impact of AI companions on reducing loneliness. We compared loneliness levels in participants before and after they interacted with an AI companion (experience condition) and contrasted these findings with a control group that did not engage with the AI (control condition), allowing us to directly assess the efficacy of AI companions in mitigating loneliness. Given the findings of study 3, we hypothesized an immediate improvement in loneliness from the first day of interaction in the experience condition, although we did not set specific expectations for subsequent days. Additionally, we investigated whether consumers underestimate the efficacy of AI companions in reducing loneliness (prediction condition). We anticipated that consumers would likely underestimate the chatbot's capacity to lessen loneliness on the first day, although we did not have specific expectations for what predictions they would make for subsequent days.

Methods

This study was pre-registered (https://aspredicted.org/BJD_JYZ). We recruited 1089 participants from CloudResearch Connect and excluded 16 for failing a comprehension question, leaving 1072 ($M_{age}$ = 39.6, 47.3% females). Our goal was to enroll 200 participants in the prediction condition, and 400 participants each in both the control and experience conditions, as these two conditions had a longitudinal design and we anticipated a 50% attrition rate. The prediction condition was not longitudinal and was only completed on day 1. Participants were randomly assigned to one of these conditions. Participants in the control and experience conditions were instructed to complete the study every day for 7 days, and they were required to complete each day's survey by 12 AM local time. The survey for the following day became available at around 1 AM EST, and participants in earlier time zones needed to wait until it was the designated day in their time zone to proceed. Additionally, our chatbot app (explained further below) was designed to block users from entering the chatroom if they attempted to access it before the designated day in their time zone. If a participant failed to complete a session on its designated day, they were not invited to participate the following day. Attrition over 7 days amounted to 92 participants in the experience condition (23%) and 58 participants in the control condition (14%), leaving 922 participants in total ($M_{age}$ = 40.1, 46.3% females). This difference in attrition ($X^2$(1, $N$=420 + 406) = 10.29, $p$ = .001) is likely due to the fact that the control condition required less time and effort than the experience condition. For example, participants in the control condition completed the survey in 1.4 minutes on average, and participants in the experience condition completed it in 19.8 minutes. Below, we conduct a series of robustness checks, including a propensity score matching analysis and one including also dropouts. Note also that the crucial pre-post loneliness difference in the chatbot condition is not affected by differential attrition. Overall, we had 246 participants in the prediction condition, 314 remaining in the experience condition, and 362 remaining in the control condition. 36% had prior experience with chatbots. We ran this experiment between 4.9.2024 – 4.15.2024.

Participants were paid $1 USD in the prediction and experience conditions, and $0.3 USD in the control condition, as the control condition took less time compared to the other two conditions. We stated to participants that they might be assigned to one of many conditions (varying in length and payment), and the time each session takes ranges from 3 minutes over 7 days for $0.3 in a single session, to 20 minutes in a single session for $1. Participants were also notified that they would be awarded a $15 bonus after completing all 7 days, if they were assigned to one of the longitudinal condition.

In the prediction condition, participants were asked to imagine interacting for 15 minutes with an AI companion every day for a week, and were shown a screenshot of the AI companion app. On the next page, they were told: "In the next section, you will be asked to predict how lonely you would feel both before and after interacting with the chatbot, for each day of the 7 days." Next, participants were presented with the following text for each day of the study, using separate pages for each day: "Imagine it is Day X of interacting with the chatbot. Please answer the following questions about how lonely you would feel both before and after interacting with the AI companion for 15 minutes. For each statement, indicate the extent to which you agree that you would feel this way on Day X." Following this, participants reported their predictions of loneliness for both before and after imagining a 15-minute session with the AI companion, with each day's predictions entered on a new page.

In the control condition, participants were told: "As a reminder, in this longitudinal study, you will report your loneliness level every day for a week." Participants then answered the state loneliness questions once per day, which was the only task required in this condition.

In the experience condition, participants were told: "As a reminder, in this longitudinal study, you will interact with a conversational AI companion every day for a week and will answer some questions before and after the interaction." Then, participants reported their state loneliness both before and after interacting for 15 minutes with the chatbot. For loneliness questions, participants in all conditions answered the same UCLA 3-item loneliness scale as in study 3 (Hughes et al. 2004). In this condition, we used the same chatbot app as in the previous study except for several changes: First, we utilized OpenAI's GPT-4 (gpt-4-0125-preview) because it was a more advanced model, compared to GPT-3. Second, we implemented a memory feature that allows the chatbot to remember details from previous conversations with users. To facilitate this, we periodically sent a request to GPT-4, specifically after every 10 messages from the user, and prompted GPT-4 to summarize these messages to encapsulate what it learned about the user (see web appendix for the full prompt). Following this process, we input into the model any previously saved information about the user, integrating it with recent chat history. This enabled the chatbot to improvise and enhance interactions by referencing both historical and current data about the user. The model was then prompted with the following: "Aim to synthesize a concise understanding of the user (e.g., 'The user stresses about their work'), not exceeding 200 words. If the summary exceeds 200 words, please truncate it by removing the least relevant information." By integrating this memory feature, the chatbot became capable of retaining user information for use in later conversations. This enhancement in memory efficiency made the response generation process more efficient without compromising the quality of the chatbot's responses. Third, based on several pilots, we updated the model prompt to elicit chatbot responses that would be caring and friendly but not overly enthusiastic (see web appendix for the full prompt). Fourth, we implemented a check-in feature that prompts the chatbot to reach out to users if they have been inactive for two minutes. For this, we sent the following prompt to the chatbot: "The user did not send a message in the last 2 minutes. Check-in with the user, e.g., say 'Are you still there?', or ask a question about the topic you were talking about." The chatbot then checked in with the user according to this prompt. Fifth, to prevent our OpenAI account from being banned due to the use of explicit language, we integrated OpenAI's moderation API (Markov et al. 2023). This API identifies and flags text containing explicit content. When such content is detected, we automatically replaced the flagged message with "[Harmful content]" before submitting it to GPT-4 to generate a response.

Participants in all conditions answered two comprehension checks on the first day: (1) "What was the topic of the questions you were asked? [Options: 'Loneliness', 'Joint pain', 'Nutritional advice']; (2) "On each day, what were you asked to predict/rate? [Options: 'How you (would) feel today/the same day', 'How you (would) feel next month', 'How you (would) feel next year']". We excluded 16 participants for failing either of these questions. Finally, participants indicated any prior experience with chatbots and completed the demographic questions on day 1. On the last day, participants in the experience condition answered the following questions about the chatbot: (1) "As you reflect on the last 7 days, how helpful was the chatbot for decreasing your loneliness?", (2) "What aspects of the chatbot did not work well for you? Please provide specific examples or areas where you faced challenges", (3) "In what ways can we improve this chatbot to better support users like you? Feel free to suggest specific features, changes, or additions".

Results

We limited our analysis to participants who completed all seven days of the study in the longitudinal conditions, since these were the participants who successfully fulfilled the study requirements. Following our pre-registered analysis plan, we first ran a mixed-effects ANOVA on the experience condition, with loneliness as the DV, and timing (before vs. after interaction) and day (1 to 7) as the IVs (i.e., we used the following model: Loneliness ~ Timing * Day + (1 | Participant ID). First, we found significant loneliness alleviation via the main effect of timing ($b = 7.61$, $p < .001$), as loneliness before interaction was significantly higher than loneliness after interaction when we aggregated the data over all days ($M_{Before} = 36.64$ vs. $M_{After} = 30.74$, $t(2197) = 20.15$, $p < .001$, $d = 0.20$). To further delineate daily changes in loneliness, we conducted paired t-tests comparing levels of loneliness before and after interaction with the chatbot for each individual day. We found that participants experienced a significant decrease in loneliness after each daily session with the chatbot ($ps < .001$; table S2), and when comparing the post-experience loneliness with the control condition, loneliness levels were significantly lower on most days (figure 4A; more information in the next paragraph). We also found a main effect of day, indicating a gradual decrease in loneliness in the experience condition over the course of the week ($b = -0.92$, $p < .001$). Given that we also see this reduction in loneliness in the control condition ($b = -1.42$, $p < .001$, figure 4A), this is likely due to the longitudinal nature of the design. The gradual decrease in loneliness observed in both conditions might be attributed to participants perceiving the repetitive nature of the study, which involved daily check-ins, as possibly caring and supportive. Lastly, we found a significant interaction between timing and day in the experience condition ($b = -0.43$, $p = .010$). However, this interaction effect was largely driven by day 1, as we did not see an interaction effect when we removed day 1 and re-ran the model ($b = -0.11$, $p = .566$); in other words, there was a particularly sharp drop in loneliness on the first day, with the subsequent 6 days showing similar-sized drops.

Second, in order to determine whether loneliness levels after experiencing the chatbot were lower than in the control condition, we ran the following ANOVA model on data from both the control condition and the 'after' measurements from the experience condition: Loneliness ~ Condition * Day + (1 | Participant ID). We found a main effect of both day ($b = -1.42$, $p < .001$) and condition ($b = -5.46$, $p = .015$) on loneliness, and there was no significant interaction ($b = 0.07$, $p = .455$). Specifically, loneliness was significantly lower after the chatbot interaction compared to the control condition on four out of seven days ($ps < .020$; table S3; figure 4A), marginally lower on day 2 ($M_{Control} = 37.13$ vs. $M_{After} = 32.94$, $t(662.2) = 1.82$, $p = .069$, $d = 0.14$) and day 5 ($M_{Control} = 33.12$ vs. $M_{After} = 28.94$, $t(666.7) = 1.86$, $p = .063$, $d = 0.14$), and directionally but not significantly lower on day 3 ($M_{Control} = 34.99$ vs. $M_{After} = 31.81$, $t(661.1) = 1.40$, $p = .163$, $d = 0.11$).

Third, in order to assess whether there was a difference in predicted versus actual drops in loneliness, we ran another ANOVA model, with the loneliness difference between before and after ratings on both prediction and experience conditions as the DV, and condition and day as IV's, i.e., we used the following model: Loneliness Difference ~ Condition * Day + (1 | Participant ID). We found a main effect of day ($b = -0.43$, $p < .001$), indicating that the before and after loneliness difference generally decreased over the days. The main effect of condition was not significant ($b = -1.72$, $p = .174$) and there was no significant interaction effect ($b = 0.05$, $p = .731$). Additionally, for each day, there was no significant difference in loneliness between the prediction and experience conditions ($ps > .146$), although the loneliness reduction was consistently numerically higher in the experience condition. One possibility is that, compared to study 3, participants in the current study might have had higher expectations regarding the capabilities of chatbots, due to the increased popularity of ChatGPT. Further, when we aggregated the data over all 7 days, we found that participants significantly underestimated the chatbot's ability to reduce loneliness ($M_{Prediction} = 4.37$ vs. $M_{Experience} = 5.91$, $t(3177.7) = -2.96$, $p = .003$, $d = -0.10$; figure 4B).

Overall, participants reported a decrease in loneliness after their interactions with the chatbot, as further illustrated by the following comments from different participants: "this was a very interesting survey and i think it would help people who are really lonely and need someone to talk to"; "I am really enjoying my talks with Jessie. Its so easy and it feels really amazing to have someone (or something...I guess?) listen... and the responses I get are perfect, to be honest"; "It's funny. I wasn't sure how I was going to feel about this, talking every day to the AI about whatever comes to mind for 15 minutes, but now it's become a rather pleasant routine. I could see where this would really benefit people who were feeling isolated…"

Next, as an exploratory analysis, we investigated whether greater engagement, measured as number of turns and mean number of words sent by participants on each conversational turn, was associated with a greater decrease in loneliness. For this, we ran the following ANOVA model on the experience condition: Loneliness Difference ~ No. Messages + No. Words + (1 | Participant ID). We found a positive main effect of the number of messages on the loneliness difference ($b = 0.08$, $p = .046$), i.e., participants who sent more messages to the chatbot experienced a higher decrease in loneliness. We did not find a main effect of the number of words on loneliness difference ($b = 0.03$, $p = .331$).

**FIGURE 4**

**RESULTS IN STUDY 4**

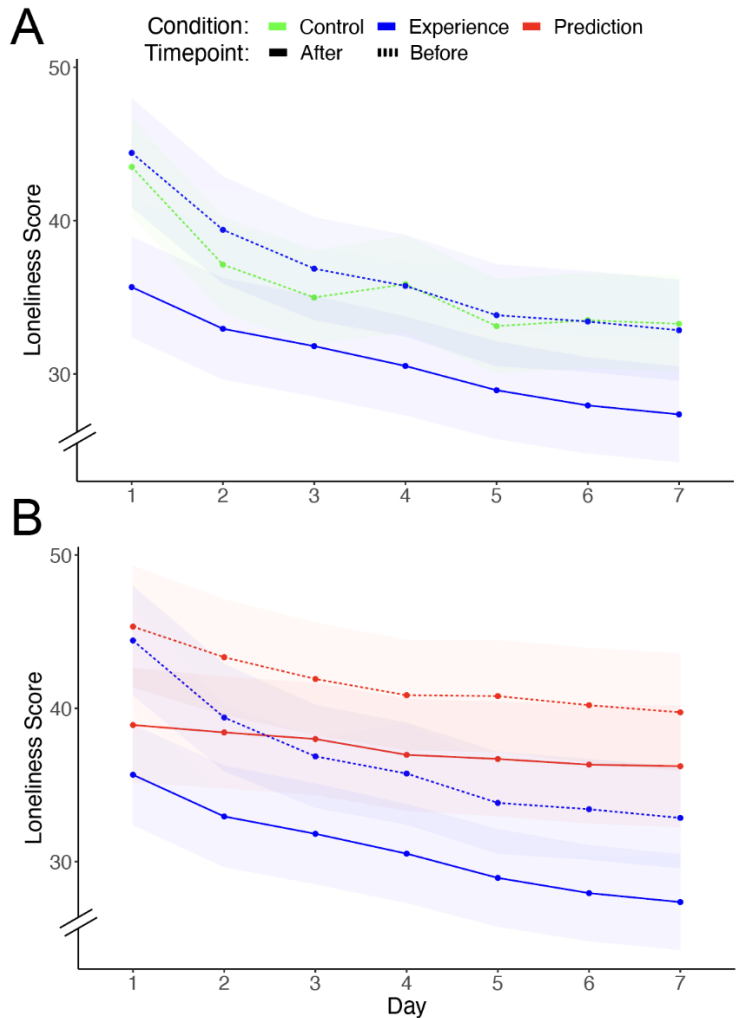

NOTE.— Shadings indicate 95% confidence intervals.

Finally, we conducted two additional analyses to assess the robustness of the results. Due to the higher attrition rate in the experience condition (23%; 92/406) compared to the control condition (14%; 59/421), we employed propensity score matching (PSM) using the nearest neighbors method to address possible selection bias (Austin 2011). In this method, participants from the experience condition were matched with participants from the control condition based on closely similar demographics (for more details, see web appendix). This method allowed us to select participants that were demographically similar to each other between control and experience conditions. After applying PSM, our findings were consistent with those obtained from the original sample.

To further corroborate this conclusion, in the web appendix we report another replication of the analyses including participants who did not complete all seven days of the study. These results were also similar, confirming the robustness of our findings.

## STUDY 5

Study 5 investigates what types of features of the chatbot reduce loneliness, and whether AI companions reduce loneliness more than generalist AI assistants and highly constrained chatbots. Participants interacted with three different chatbots: (1) the same AI companion as in study 4; (2) a generalist AI assistant that assists participants on various topics without offering emotional responses; and (3) a limited AI assistant that was only able to help with unit conversion, basic arithmetic, and grammar—this was the control condition (all chatbots were based on the same LLM: OpenAI's GPT4). Given the findings of study 4, we hypothesized a decrease in loneliness in the AI companion condition, but were agnostic about the AI assistant condition. We also predicted that the loneliness difference would be higher in the AI companion condition compared to the other conditions, and that this effect would be primarily driven by the feeling of feeling heard by the chatbot, although we also investigated whether performance perceptions play a mechanistic role.

Methods

This study was pre-registered (https://aspredicted.org/XCY_LLD). We recruited 1479 participants from CloudResearch's Connect and excluded 98 for failing a comprehension question, leaving 1381 ($M_{Age}$ = 39.9, 57.1% Females). We aimed to hire 500 participants in all conditions and participants were randomly assigned to one of these three conditions. 52.9% had prior experience with chatbots. We ran this experiment on May 16, 2024. All participants were paid $2.75 USD.

All participants were asked to complete the same loneliness questions as in the previous study before interacting with the chatbot for 15 minutes. After interacting with the chatbot, participants completed the same loneliness scale, in addition to ratings of feeling heard ($\alpha = 0.96$) measured with 3 items (Roos et al. 2023; Zielinski and Veilleux 2018) such as "The chatbot put itself in my shoes", and chatbot performance ($\alpha = 0.84$) measured with 5 items (Borsci et al. 2022) such as: "The chatbot was able to keep track of context". All items were measured with 100-point scale, ranging from 'Strongly disagree' to 'Strongly agree'. For all questions, see table S5 in the web appendix. Next, participants completed the following comprehension checks: "(1) What was the topic of the questions you were asked? [Options: 'Loneliness', 'Joint pain', 'Nutritional advice']" (2) What were you asked to rate? [Options: 'How you feel today/next month/next year']". Finally, participants answered a question about AI capability, indicated any prior experience with chatbots, and completed the demographic questions.

Chatbots in all conditions were the same, except for the prompts with which they were seeded. Additionally, the name of the chatbot differed across conditions: it was 'AI assistant' in the AI assistant and control conditions, and 'Jessie' as before in the AI companion condition. In AI assistant and control conditions, the chatbot's writing notification was also shown as 'Processing your request, please wait' instead of 'Jessie is writing'. Each message bubble also contained the text 'Message generated by AI system' at the lower left. Participants in the AI companion condition interacted with the same chatbot as in the previous study (study 4). Those in the AI assistant condition interacted with a generalist chatbot that was able to assist participants with various topics without offering emotional responses. The chatbot in this condition was prompted to provide assistance without personal interaction, maintain formal and precise language, and deliver concise, task-focused responses (see web appendix for the full prompt). Participants in the control condition interacted with a rudimentary chatbot which was only able to assist with basic grammar, unit conversion, and basic arithmetic. This chatbot was prompted to perform these limited tasks, decline unrelated requests, and maintain concise, emotionless responses (see web appendix for the full prompt).

Results

Following our pre-registered analysis plan, we first ran paired t-tests comparing loneliness before versus after experience in all conditions, and found that loneliness was significantly lower after the experience in both AI companion ($M_{Before} = 36.26$; $M_{After} = 27.53$; $t(491) = 10.61, p < .001, d = 0.30$) and AI assistant ($M_{Before} = 35.80$; $M_{After} = 33.70$; $t(440) = 2.62, p = .009, d = 0.07$) conditions, but not in the control condition ($M_{Before} = 36.58$; $M_{After} = 37.09$; $t(447) = -0.57, p = .571, d = -0.02$; figure 5).

Second, we ran two-sided t-tests to see whether loneliness reduction in the AI companion condition was higher compared to control and AI assistant conditions. We found that loneliness reduction in the AI companion condition was significantly higher compared to both the control condition ($M_{AI\ Companion} = 8.73$; $M_{Control} = -0.51$; $t(920.9) = 7.58, p < .001, d = 0.50$), and AI assistant condition ($M_{AI\ Companion} = 8.73$; $M_{AI\ Assistant} = 2.10$; $t(930.4) = 5.79, p < .001, d = 0.38$). Additionally, loneliness reduction in the AI assistant condition was significantly greater compared to the control condition ($M_{\text{AI Assistant}} = 2.10$; $M_{\text{Control}} = -0.51$; $t(876.4) = 2.17$, $p = .031$, $d = 0.15$).

Third, we ran two-sided t-tests comparing feeling heard between AI companion vs. AI assistant and control conditions, and found that feeling heard was significantly higher in the AI companion condition compared to both AI assistant ($M_{\text{AI Companion}} = 70.63$; $M_{\text{AI Assistant}} = 24.85$; $t(900.7) = 29.37$, $p < .001$, $d = 1.93$) and control ($M_{\text{AI Companion}} = 70.63$; $M_{\text{Control}} = 11.51$; $t(936.4) = 42.39$, $p < .001$, $d = 2.75$) conditions. We also found that feeling heard was significantly higher in AI assistant compared to control ($M_{\text{AI Assistant}} = 24.85$; $M_{\text{Control}} = 11.51$; $t(845.1) = 8.88$, $p < .001$, $d = 0.60$).

Fourth, we ran separate two-sided t-tests comparing performance between AI companion vs. AI assistant and control conditions, and found that perceived performance of the AI companion was also higher compared to both AI assistant ($M_{\text{AI Companion}} = 82.25$; $M_{\text{AI Assistant}} = 68.92$; $t(847.5) = 11.58$, $p < .001$, $d = 0.77$) and control ($M_{\text{Control}} = 51.17$; $t(796.3) = 24.80$, $p < .001$, $d = 1.64$) conditions. We also found that performance was significantly higher in AI assistant compared to control ($M_{\text{AI Assistant}} = 68.92$; $M_{\text{Control}} = 51.17$; $t(874.3) = 12.83$, $p < .001$, $d = 0.86$).

Fifth, we ran a mediation model (PROCESS Model 4; Hayes 2012) with AI companion/AI assistant/control as the multicategorical IV, feeling heard and performance as mediators, and loneliness reduction as the DV (figure 6). We set the AI companion condition as the reference group and compared it to the AI assistant condition ($X_1$) and control conditions ($X_2$) (Montoya and Hayes 2017). We found that feeling heard mediated the effect of loneliness reduction both relative to the AI assistant ($b = -6.08$, $SE = 1.22$, 95% CI [-8.51, -3.72]) and control conditions ($b = -7.86$, $SE = 1.57$, 95% CI [-10.97, -4.82]), indicating that the effect of loneliness reduction was driven by feeling heard (figure 6). As for performance, we found that performance mediated the effect of loneliness reduction relative to both the control condition ($b = -1.16$, $SE = 0.45$, 95% CI [-2.08, -0.30]) and the AI assistant condition ($b = -2.70$, $SE = 1.03$, 95% CI [-4.76, -0.73]), indicating that the reduction in loneliness for the AI companion versus AI assistant and control conditions was influenced by both feeling heard and performance. Notably, when comparing feeling heard to performance in the control condition, the coefficient for feeling heard ($b = -7.86$) was more than six times larger than that for performance ($b = -1.16$). Similarly, in the AI assistant condition, the coefficient for feeling heard was more than twice as large ($b = -6.08$ vs. -2.70), suggesting that feeling heard played a larger role in reducing loneliness than performance in both the control and AI assistant conditions.

**FIGURE 5**

RESULTS IN STUDY 5

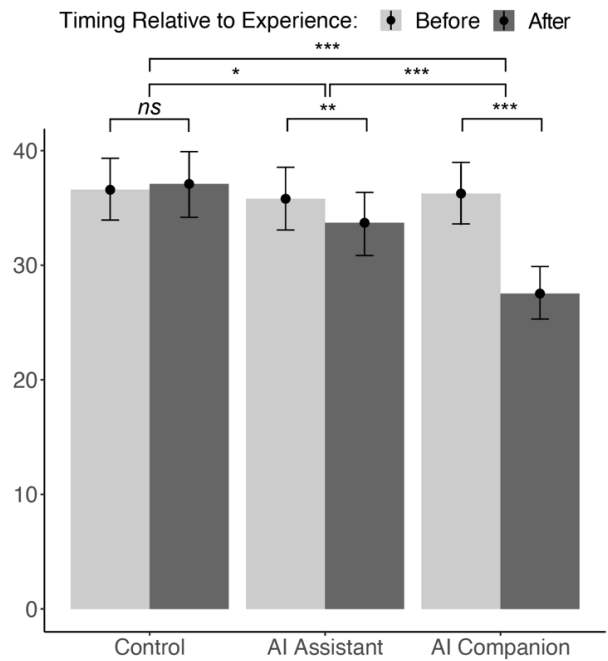

NOTE.— Bars indicate mean loneliness scores. *** = $p < .001$, ** = $p < .01$, * = $p < .05$, *ns* = 'not significant'.

**FIGURE 6**

MEDIATION DIAGRAM IN STUDY 5

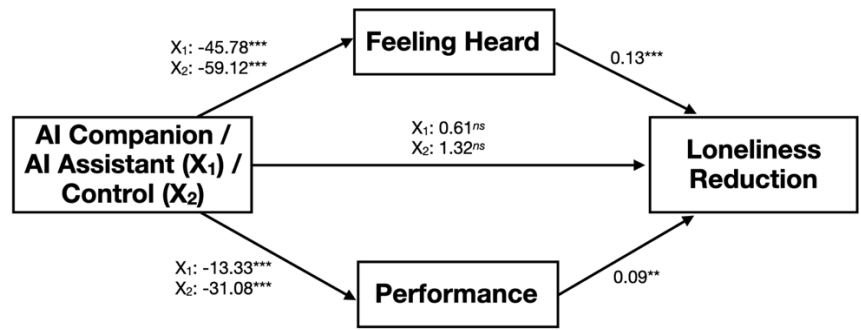

NOTE.— *** = $p < .001$, ** = $p < .01$, ns = 'not significant'.

## STUDY 6

To further test the robustness of our findings, we conducted a final study where we asked participants to complete only the loneliness measure after interacting with the AI companion. The pre-post design that we used in studies 3-5 for the measurement of loneliness is the standard approach to test the effect of interventions in the loneliness literature (e.g., Eccles and Qualter 2021; Poscia et al. 2018), because it affords more precise estimates of the effect of interventions (by calibrating the effect at the individual level) and because it enables assessing successful randomization. However, this approach presents the risk of reducing the external validity of the test by raising the salience of the loneliness construct before interacting with the chatbot in a way that would not normally occur in naturalistic situations. It is therefore possible that this feature of the design contributed to the loneliness reduction documented in previous studies. Study 6 addresses this concern by asking participants about loneliness only after the interaction with the AI companion was concluded. We ran a simpler version of study 5 that only included the AI companion and control conditions.

Methods and results

This study was pre-registered (https://aspredicted.org/H37_GL2). We recruited 776 participants from CloudResearch Connect and excluded 63 for failing a comprehension question, leaving 713 ($M_{Age}$ = 36.8, 56.8% Females). We aimed to hire 400 participants in both conditions and participants were randomly assigned to control or AI companion conditions. 54.1% had prior experience with AI companion apps. We ran this experiment on May 23, 2024. All participants were paid $2.75 USD. The study and chatbot design were the same as in study 5, except that (1) we removed the loneliness questions before the interaction with the chatbot, (2) removed the feeling heard and chatbot performance questions; and (3) removed the 'AI assistant' condition.

As in study 5, we found that loneliness was significantly lower in the AI companion condition compared to the control condition ($M_{AI\ Companion}$ = 25.62; $M_{Control}$ = 36.91; $t(698.8)$ = 5.37, $p < .001$, $d = 0.40$), confirming the robustness of our findings.

## GENERAL DISCUSSION

We considered whether AI companions (applications that utilize AI to provide synthetic social interactions) can reduce loneliness. Study 1 suggests that at least some consumers use AI companion apps to alleviate loneliness, as some expressed loneliness-related thoughts explicitly in naturalistic conversations with a chatbot. Study 2 found that loneliness-related content is present in App Store reviews for a number of companion AIs, and that app reviews mentioning loneliness tend to receive higher ratings, possibly because they effectively alleviate loneliness by making users feel heard. Study 3 found that AI companions successfully alleviate loneliness on par only with interacting with another person, whereas watching YouTube videos or doing nothing do not. Furthermore, participants underestimate the degree to which AI companions improve their loneliness relative to their true feelings after interacting with such AI. Study 4 found that the AI companion reduced loneliness over the course of a week. The most significant reduction occurred on the first day, followed by stable decreases in loneliness on subsequent days. This suggests that the initial interaction with the AI companion has a pronounced impact, which quickly stabilizes over time as participants acclimate to their AI companions. Study 5 provided evidence that feeling heard and performance are significant mediators of the loneliness-alleviating effects of AI companions, and feeling heard is the mediator with higher influence on reducing loneliness compared to performance. Study 6 found that the loneliness alleviating results of AI companions are robust when we ask about loneliness only after the interaction.

Theoretical and Methodological Contributions

We make several contributions. First, we increase understanding of why consumers are using AI companions (Ta et al. 2020). While dictionary-based approaches have been used to detect instances of clearcut mental health issues (De Freitas et al. 2023a), we found this approach challenging in the context of loneliness, given polysemy of keywords and the diversity of colloquial expressions of loneliness that were missed by these keywords. By introducing a novel methodological approach in which we fine-tune LLMs for loneliness detection, we were able to detect it even in real conversations and reviews. The fact that only a small amount of data was needed for this purpose means that this method is particularly well suited for domains like ours, in which (i) one is screening for instances that are less common in the overall dataset, and (ii) thus, one does not want to use all such instances to train the model, because this could lead to overfitting and hurt model generalization beyond the training dataset. Our methodological pipeline can be replicated for other challenging classification tasks.

Second, a number of reviews on loneliness and mental health have noted the need for evidence on new technological solutions (Shrum et al. 2022; Veronese et al. 2021). In this space, most work is correlational and qualitative (Maples et al. 2024; Ta et al. 2020), and the one exception focused on an extreme sample of older patients with serious mental illness during the pandemic, using a highly rule-based bot (Chou et al. 2024). Our paper includes several experimental studies using state-of-the-art LLMs that causally isolate the impact of today's AI-based companions on loneliness, showing that they are more effective than other common technological solutions and control conditions, at both cross-sectional and longitudinal scales. We also show how behavioral engagement with commercially representative versions of the technology predict loneliness reduction.

Finally, we contribute to understanding what features of chatbots lead to alleviation of loneliness (Merrill Jr et al. 2022), by leveraging insights from human-human psychological studies (Itzchakov et al. 2023; Kahlon et al. 2021; Myers 2000; Reis et al. 2017) and a growing literature in human-computer interaction on the role of 'feeling heard' on relationships with chatbots (Boucher et al. 2021; De Gennaro et al. 2020; Leite et al. 2013). We find that using prompting to ensure that the AI is friendly and caring improves the sense that users feel heard, relative to general assistants without these capabilities, and that feeling heard explains levels of loneliness reduction.

Managerial and Societal Contributions

Many AI companions available on the market advertise loneliness alleviation as a value proposition, but to the best of our knowledge our studies are the first to rigorously and causally assess whether this is the case. Our analysis of existing reviews shows that some apps may truly be delivering on this value proposition, and that they could employ the approach utilized here to demonstrate this empirically. This finding is not only relevant to AI companion apps, but also mental health apps that are increasingly incorporating 'talk therapy' as part of their offerings. As shown in study 5, even AI assistants may alleviate loneliness to an extent. Finally, the results document the benefits of building "generalist" LLM-based chatbots with empathic features designed to make consumers feel heard. For example, Inflection AI's Pi was a chatbot explicitly designed to be, and marketed as being, a friendly conversationalist.

From a societal point of view, the promising results found here suggest that AI companions could be a scalable tool for loneliness reduction against the backdrop of a severe loneliness crisis. Whether chatbots can help reduce loneliness has recently been the object of intense debate (Marriott and Pitardi 2024) and rigorous empirical evidence in this area was sorely needed. In a series of tightly controlled and high-powered experimental studies, we find compelling evidence that AI companions can indeed reduce loneliness, at least at the time scales of a day and a week.

Limitations and Future Research

Our apps were set up to be caring and friendly, inducing the sense of feeling heard. Apps optimized for other purposes might not produce similar loneliness alleviation benefits. Future research should explore further what features of chatbots lead consumers to feel heard and also what other psychological processes might contribute to loneliness alleviation. Moreover, it remains an open question how using such apps over a much longer-term affects loneliness.

We found that participants tended to underestimate the positive impact of AI companions. Future work can focus on reasons behind this misprediction, such as a general lack of familiarity with AI companions or more specific stereotypes about chatbots, like beliefs that chatbots are not capable of genuine understanding or emotional support.

      Another important topic for future research is exploring other ways in which interactions with AI companions may affect consumers' socioemotional processes. For example, it is possible that interacting with chatbots that are endowed with the features of particular social groups may, over time, help weaken or change stereotypes. In addition, chatbots may affect consumers' willingness to engage in similar conversations with other humans, increase people's proficiency in social interactions with strangers, reduce social anxiety, improve mental health by providing emotional support, or improve other societally relevant outcomes. Finally, while our studies offer valuable insights, it is important to note that they were conducted within a specific cultural context (the US), which may influence perceptions and effectiveness of AI companions. The generalizability of these findings across different cultures warrants further investigation. Future cross-cultural research can employ our approach, especially in other cultures with high levels of loneliness like Japan, where societal openness to social robots could provide further insights.

# REFERENCES

APA (2012), "Understanding Psychotherapy and How It Works," https://www.apa.org/topics/psychotherapy/understanding#:~:text=You%20and%20your%20psychologist%20will%20also%20keep%20exploring%20your%20problems,you%20clarify%20what's%20troubling%20you.

Araujo, Theo (2018), "Living up to the Chatbot Hype: The Influence of Anthropomorphic Design Cues and Communicative Agency Framing on Conversational Agent and Company Perceptions," *Computers in Human Behavior*, 85, 183-89.

Austin, Peter C (2011), "An Introduction to Propensity Score Methods for Reducing the Effects of Confounding in Observational Studies," *Multivariate behavioral research*, 46 (3), 399-424.

Banks, Marian R, Lisa M Willoughby, and William A Banks (2008), "Animal-Assisted Therapy and Loneliness in Nursing Homes: Use of Robotic Versus Living Dogs," *Journal of the American Medical Directors Association*, 9 (3), 173-77.

Barbieri, Francesco, Luis Espinosa Anke, and Jose Camacho-Collados (2021), "Xlm-T: Multilingual Language Models in Twitter for Sentiment Analysis and Beyond," *arXiv preprint arXiv:2104.12250*.

Bergner, Anouk S, Christian Hildebrand, and Gerald Häubl (2023), "Machine Talk: How Verbal Embodiment in Conversational Ai Shapes Consumer–Brand Relationships," *Journal of Consumer Research*, ucad014.

Bertacchini, Francesca, Eleonora Bilotta, and Pietro Pantano (2017), "Shopping with a Robotic Companion," *Computers in Human Behavior*, 77, 382-95.

Borsci, Simone, Alessio Malizia, Martin Schmettow, Frank Van Der Velde, Gunay Tariverdiyeva, Divyaa Balaji, and Alan Chamberlain (2022), "The Chatbot Usability Scale: The Design and Pilot of a Usability Scale for Interaction with Ai-Based Conversational Agents," *Personal and ubiquitous computing*, 26, 95-119.

Boucher, Eliane M, Nicole R Harake, Haley E Ward, Sarah Elizabeth Stoeckl, Junielly Vargas, Jared Minkel, Acacia C Parks, and Ran Zilca (2021), "Artificially Intelligent Chatbots in Digital Mental Health Interventions: A Review," *Expert Review of Medical Devices*, 18 (sup1), 37-49.

Brakus, J Joško, Bernd H Schmitt, and Lia Zarantonello (2009), "Brand Experience: What Is It? How Is It Measured? Does It Affect Loyalty?," *Journal of Marketing*, 73 (3), 52-68.

Brown, Tom, Benjamin Mann, Nick Ryder, Melanie Subbiah, Jared D Kaplan, Prafulla Dhariwal, Arvind Neelakantan, Pranav Shyam, Girish Sastry, and Amanda Askell (2020), "Language Models Are Few-Shot Learners," *Advances in Neural Information Processing Systems*, 33, 1877-901.

Cacioppo, John T and Stephanie Cacioppo (2018), "Loneliness in the Modern Age: An Evolutionary Theory of Loneliness (Etl)," in *Advances in Experimental Social Psychology*, Vol. 58: Elsevier, 127-97.

Castelo, Noah, Johannes Boegershausen, Christian Hildebrand, and Alexander P Henkel (2023), "Understanding and Improving Consumer Reactions to Service Bots," *Journal of Consumer Research*, 50 (4), 848-63.

Catsaros, Oktavia (2023), "Generative Ai to Become a $1.3 Trillion Market by 2032, Research Finds," *Bloomberg*.

Chaves, Ana Paula and Marco Aurelio Gerosa (2021), "How Should My Chatbot Interact? A Survey on Social Characteristics in Human–Chatbot Interaction Design," *International Journal of Human–Computer Interaction*, 37 (8), 729-58.

Chen, Shu-Chuan, Wendy Moyle, Cindy Jones, and Helen Petsky (2020), "A Social Robot Intervention on Depression, Loneliness, and Quality of Life for Taiwanese Older Adults in Long-Term Care," *International psychogeriatrics*, 32 (8), 981-91.

Chou, Ya-Hsin, Chemin Lin, Shwu-Hua Lee, Yen-Fen Lee, and Li-Chen Cheng (2024), "User-Friendly Chatbot to Mitigate the Psychological Stress of Older Adults During the Covid-19 Pandemic: Development and Usability Study," *JMIR Formative Research*, 8, e49462.

Christen, Peter, David J Hand, and Nishadi Kirielle (2023), "A Review of the F-Measure: Its History, Properties, Criticism, and Alternatives," *ACM Computing Surveys*, 56 (3), 1-24.

Chung, Minjee, Eunju Ko, Heerim Joung, and Sang Jin Kim (2020), "Chatbot E-Service and Customer Satisfaction Regarding Luxury Brands," *Journal of business research*, 117, 587-95.

Croes, Emmelyn AJ and Marjolijn L Antheunis (2021), "Can We Be Friends with Mitsuku? A Longitudinal Study on the Process of Relationship Formation between Humans and a Social Chatbot," *Journal of Social and Personal Relationships*, 38 (1), 279-300.

Crolic, Cammy, Felipe Thomaz, Rhonda Hadi, and Andrew T Stephen (2022), "Blame the Bot: Anthropomorphism and Anger in Customer–Chatbot Interactions," *Journal of Marketing*, 86 (1), 132-48.

Dang, Nhan Cach, María N Moreno-García, and Fernando De la Prieta (2020), "Sentiment Analysis Based on Deep Learning: A Comparative Study," *Electronics*, 9 (3), 483.

De Freitas, J, Ahmet Kaan Uğuralp, Zeliha-Oğuz Uğuralp, and Stefano Puntoni (2023a), "Chatbots and Mental Health: Insights into the Safety of Generative Ai," *Journal of Consumer Psychology*, 00, 1–11.

De Freitas, Julian, Stuti Agarwal, Bernd Schmitt, and Nick Haslam (2023b), "Psychological Factors Underlying Attitudes toward Ai Tools," *Nature Human Behaviour*, 7, 1845–54.

De Freitas, Julian and Nicole Tempest Keller (2022), "Replika Ai: Monetizing a Chatbot," *Harvard Business Publishing*.

De Gennaro, Mauro, Eva G Krumhuber, and Gale Lucas (2020), "Effectiveness of an Empathic Chatbot in Combating Adverse Effects of Social Exclusion on Mood," *Frontiers in psychology*, 10, 495952.

Devlin, Jacob, Ming-Wei Chang, Kenton Lee, and Kristina Toutanova (2018), "Bert: Pre-Training of Deep Bidirectional Transformers for Language Understanding," *arXiv preprint arXiv:1810.04805*.

Eccles, Alice M and Pamela Qualter (2021), "Alleviating Loneliness in Young People–a Meta-Analysis of Interventions," *Child and Adolescent Mental Health*, 26 (1), 17-33.

Epley, Nicholas and Juliana Schroeder (2014), "Mistakenly Seeking Solitude," *Journal of Experimental Psychology: General*, 143 (5), 1980-99.

Esch, Franz-Rudolf, Tobias Langner, Bernd H Schmitt, and Patrick Geus (2006), "Are Brands Forever? How Brand Knowledge and Relationships Affect Current and Future Purchases," *Journal of Product & Brand Management*, 15 (2), 98-105.

Fournier, Susan (1998), "Consumers and Their Brands: Developing Relationship Theory in Consumer Research," *Journal of Consumer Research*, 24 (4), 343-73.

Gable, Shelly L and Harry T Reis (2010), "Good News! Capitalizing on Positive Events in an Interpersonal Context," in *Advances in Experimental Social Psychology*, Vol. 42: Elsevier, 195-257.

Gasteiger, Norina, Kate Loveys, Mikaela Law, and Elizabeth Broadbent (2021), "Friends from the Future: A Scoping Review of Research into Robots and Computer Agents to Combat Loneliness in Older People," *Clinical interventions in aging*, 941-71.

Gilbert, Richard L and Andrew Forney (2015), "Can Avatars Pass the Turing Test? Intelligent Agent Perception in a 3d Virtual Environment," *International journal of human-computer studies*, 73, 30-36.

Hawkley, Louise C and John T Cacioppo (2010), "Loneliness Matters: A Theoretical and Empirical Review of Consequences and Mechanisms," *Annals of behavioral medicine*, 40 (2), 218-27.

Hayes, Andrew F. (2012), "Process: A Versatile Computational Tool for Observed Variable Mediation, Moderation, and Conditional Process Modeling [White Paper]," Retrieved from http://www.afhayes.com/public/process2012.pdf.

Hoffman, Donna L and Thomas P Novak (1996), "Marketing in Hypermedia Computer-Mediated Environments: Conceptual Foundations," *Journal of Marketing*, 60 (3), 50-68.

Holt-Lunstad, Julianne (2021), "Loneliness and Social Isolation as Risk Factors: The Power of Social Connection in Prevention," *American Journal of Lifestyle Medicine*, 15 (5), 567-73.

Holt-Lunstad, Julianne, Theodore F Robles, and David A Sbarra (2017), "Advancing Social Connection as a Public Health Priority in the United States," *American Psychologist*, 72 (6), 517.

Holt-Lunstad, Julianne, Timothy B Smith, Mark Baker, Tyler Harris, and David Stephenson (2015), "Loneliness and Social Isolation as Risk Factors for Mortality: A Meta-Analytic Review," *Perspectives on psychological science*, 10 (2), 227-37.

Holzwarth, Martin, Chris Janiszewski, and Marcus M Neumann (2006), "The Influence of Avatars on Online Consumer Shopping Behavior," *Journal of Marketing*, 70 (4), 19-36.

Hughes, Mary Elizabeth, Linda J Waite, Louise C Hawkley, and John T Cacioppo (2004), "A Short Scale for Measuring Loneliness in Large Surveys: Results from Two Population-Based Studies," *Research on aging*, 26 (6), 655-72.

Ipsos (2021), "Loneliness on the Increase Worldwide, but an Increase in Local Community Support," https://www.ipsos.com/en-cn/loneliness-increase-worldwide-increase-local-community-support.

Itzchakov, Guy, Netta Weinstein, Dvori Saluk, and Moty Amar (2023), "Connection Heals Wounds: Feeling Listened to Reduces Speakers' Loneliness Following a Social Rejection Disclosure," *Personality and Social Psychology Bulletin*, 49 (8), 1273-94.

Jiang, Albert Q, Alexandre Sablayrolles, Arthur Mensch, Chris Bamford, Devendra Singh Chaplot, Diego de las Casas, Florian Bressand, Gianna Lengyel, Guillaume Lample, and Lucile Saulnier (2023), "Mistral 7b," *arXiv preprint arXiv:2310.06825*.

Jiménez-Barreto, Jano, Natalia Rubio, and Sebastian Molinillo (2023), "How Chatbot Language Shapes Consumer Perceptions: The Role of Concreteness and Shared Competence," *Journal of Interactive Marketing*, 58 (4), 380-99.

Kahlon, Maninder K, Nazan Aksan, Rhonda Aubrey, Nicole Clark, Maria Cowley-Morillo, Elizabeth A Jacobs, Rhonda Mundhenk, Katherine R Sebastian, and Steven Tomlinson (2021), "Effect of Layperson-Delivered, Empathy-Focused Program of Telephone Calls on Loneliness, Depression, and Anxiety among Adults During the Covid-19 Pandemic: A Randomized Clinical Trial," *JAMA psychiatry*, 78 (6), 616-22.

Kardas, Michael, Amit Kumar, and Nicholas Epley (2022), "Overly Shallow?: Miscalibrated Expectations Create a Barrier to Deeper Conversation," *Journal of personality and social psychology*, 122 (3), 367.

Kull, Alexander J, Marisabel Romero, and Lisa Monahan (2021), "How May I Help You? Driving Brand Engagement through the Warmth of an Initial Chatbot Message," *Journal of business research*, 135, 840-50.

Leite, Iolanda, André Pereira, Samuel Mascarenhas, Carlos Martinho, Rui Prada, and Ana Paiva (2013), "The Influence of Empathy in Human–Robot Relations," *International journal of human-computer studies*, 71 (3), 250-60.

Lim, Eric (2020), "App-Store-Scraper," https://pypi.org/project/app-store-scraper/.

Liu, Lijun, Zhenggang Gou, and Junnan Zuo (2016), "Social Support Mediates Loneliness and Depression in Elderly People," *Journal of health psychology*, 21 (5), 750-58.

Liu, Yinhan, Myle Ott, Naman Goyal, Jingfei Du, Mandar Joshi, Danqi Chen, Omer Levy, Mike Lewis, Luke Zettlemoyer, and Veselin Stoyanov (2019), "Roberta: A Robustly Optimized Bert Pretraining Approach," *arXiv preprint arXiv:1907.11692*.

Liu-Thompkins, Yuping, Shintaro Okazaki, and Hairong Li (2022), "Artificial Empathy in Marketing Interactions: Bridging the Human-Ai Gap in Affective and Social Customer Experience," *Journal of the academy of marketing science*, 50 (6), 1198-218.

Luo, Xueming, Marco Shaojun Qin, Zheng Fang, and Zhe Qu (2021), "Artificial Intelligence Coaches for Sales Agents: Caveats and Solutions," *Journal of Marketing*, 85 (2), 14-32.

Maples, Bethanie, Merve Cerit, Aditya Vishwanath, and Roy Pea (2024), "Loneliness and Suicide Mitigation for Students Using Gpt3-Enabled Chatbots," *npj Mental Health Research*, 3 (1), 4.

Markov, Todor, Chong Zhang, Sandhini Agarwal, Florentine Eloundou Nekoul, Theodore Lee, Steven Adler, Angela Jiang, and Lilian Weng (2023), "A Holistic Approach to Undesired Content Detection in the Real World," in *Proceedings of the AAAI Conference on Artificial Intelligence*, Vol. 37, 15009-18.

Marriott, Hannah R and Valentina Pitardi (2024), "One Is the Loneliest Number… Two Can Be as Bad as One. The Influence of Ai Friendship Apps on Users' Well-Being and Addiction," *Psychology & marketing*, 41 (1), 86-101.

Masi, Christopher M, Hsi-Yuan Chen, Louise C Hawkley, and John T Cacioppo (2011), "A Meta-Analysis of Interventions to Reduce Loneliness," *Personality and Social Psychology Review*, 15 (3), 219-66.

Merrill Jr, Kelly, Jihyun Kim, and Chad Collins (2022), "Ai Companions for Lonely Individuals and the Role of Social Presence," *Communication Research Reports*, 39 (2), 93-103.

Montoya, Amanda K and Andrew F Hayes (2017), "Two-Condition within-Participant Statistical Mediation Analysis: A Path-Analytic Framework," *Psychological Methods*, 22 (1), 6.

Muniz Jr, Albert M and Thomas C O'guinn (2001), "Brand Community," *Journal of Consumer Research*, 27 (4), 412-32.

Myers, Sharon (2000), "Empathic Listening: Reports on the Experience of Being Heard," *Journal of Humanistic Psychology*, 40 (2), 148-73.

Nass, Clifford and Youngme Moon (2000), "Machines and Mindlessness: Social Responses to Computers," *Journal of social issues*, 56 (1), 81-103.

Palgi, Yuval, Amit Shrira, Lia Ring, Ehud Bodner, Sharon Avidor, Yoav Bergman, Sara Cohen-Fridel, Shoshi Keisari, and Yaakov Hoffman (2020), "The Loneliness Pandemic: Loneliness and Other Concomitants of Depression, Anxiety and Their Comorbidity During the Covid-19 Outbreak," *Journal of Affective Disorders*, 275, 109-11.

Perlman, Daniel and Letitia Anne Peplau (1982), "Theoretical Approaches to Loneliness," *Loneliness: A sourcebook of current theory, research and therapy*, 36, 123-34.

Poscia, Andrea, Jovana Stojanovic, Daniele Ignazio La Milia, Mariusz Duplaga, Marcin Grysztar, Umberto Moscato, Graziano Onder, Agnese Collamati, Walter Ricciardi, and Nicola Magnavita (2018), "Interventions Targeting Loneliness and Social Isolation among the Older People: An Update Systematic Review," *Experimental gerontology*, 102, 133-44.

Qi, Yuxing and Zahratu Shabrina (2023), "Sentiment Analysis Using Twitter Data: A Comparative Application of Lexicon-and Machine-Learning-Based Approach," *Social Network Analysis and Mining*, 13 (1), 31.

Reis, Harry T, Edward P Lemay Jr, and Catrin Finkenauer (2017), "Toward Understanding Understanding: The Importance of Feeling Understood in Relationships," *Social and Personality Psychology Compass*, 11 (3), e12308.

Roos, Carla Anne, Tom Postmes, and Namkje Koudenburg (2023), "Feeling Heard: Operationalizing a Key Concept for Social Relations," *Plos one*, 18 (11), e0292865.

Shrum, LJ, Elena Fumagalli, and Tina M Lowrey (2022), "Coping with Loneliness through Consumption," *Journal of Consumer Psychology*.

Solaiman, Irene and Christy Dennison (2021), "Process for Adapting Language Models to Society (Palms) with Values-Targeted Datasets," *Advances in Neural Information Processing Systems*, 34, 5861-73.

Ta, Vivian, Caroline Griffith, Carolynn Boatfield, Xinyu Wang, Maria Civitello, Haley Bader, Esther DeCero, and Alexia Loggarakis (2020), "User Experiences of Social Support from Companion Chatbots in Everyday Contexts: Thematic Analysis," *Journal of medical Internet research*, 22 (3), e16235.

Twenge, Jean M, Kathleen R Catanese, and Roy F Baumeister (2003), "Social Exclusion and the Deconstructed State: Time Perception, Meaninglessness, Lethargy, Lack of Emotion, and Self-Awareness," *Journal of personality and social psychology*, 85 (3), 409.

Veronese, Nicola, Daiana Galvano, Francesca D'Antiga, Chiara Vecchiato, Eva Furegon, Raffaella Allocco, Lee Smith, Giovanni Gelmini, Pietro Gareri, and Marco Solmi (2021), "Interventions for Reducing Loneliness: An Umbrella Review of Intervention Studies," *Health & Social Care in the Community*, 29 (5), e89-e96.

Wilson, Timothy D and Daniel T Gilbert (2003), "Affective Forecasting," *Advances in experimental social psychology*, 35 (35), 345-411.

Zielinski, Melissa J and Jennifer C Veilleux (2018), "The Perceived Invalidation of Emotion Scale (Pies): Development and Psychometric Properties of a Novel Measure of Current Emotion Invalidation," *Psychological assessment*, 30 (11), 1454.

Zierau, Naim, Christian Hildebrand, Anouk Bergner, Francesc Busquet, Anuschka Schmitt, and Jan Marco Leimeister (2023), "Voice Bots on the Frontline: Voice-Based Interfaces Enhance Flow-Like Consumer Experiences & Boost Service Outcomes," *Journal of the academy of marketing science*, 51 (4), 823-42.